# MICROSOFT DEFENDER WILL BE DEFENDED: MEMORYRANGER PREVENTS BLINDING WINDOWS AV

Denis Pogonin
Bachelor of Information Security, MEPhI
Moscow, Russia
denpog00@gmail.com

Igor Korkin, PhD
Security Researcher
Moscow, Russia
igor.korkin@gmail.com

**ABSTRACT**

Windows OS is facing a huge rise in kernel attacks. An overview of popular techniques that result in loading kernel drivers will be presented. One of the key targets of modern threats is disabling and blinding Microsoft Defender, a default Windows AV. The analysis of recent driver-based attacks will be given, the challenge is to block them. The survey of user- and kernel-level attacks on Microsoft Defender will be given. One of the recently published attackers' techniques abuses Mandatory Integrity Control (MIC) and Security Reference Monitor (SRM) by modifying Integrity Level and Debug Privileges for the Microsoft Defender via syscalls. However, this user-mode attack can be blocked via the Windows "trust labels" mechanism. The presented paper discovered the internals of MIC and SRM, including the analysis of Microsoft Defender during malware detection. We show how attackers can attack Microsoft Defender using a kernel-mode driver. This driver modifies the fields of the Token structure allocated for the Microsoft Defender application. The presented attack resulted in disabling Microsoft Defender, without terminating any of its processes and without triggering any Windows security features, such as PatchGuard. The customized hypervisor-based solution named MemoryRanger was used to protect the Windows Defender kernel structures. The experiments show that MemoryRanger successfully restricts access to the sensitive kernel data from illegal access attempts with affordable performance degradation.

**Keywords**: hypervisor-based protection, Mandatory Integrity Control, Security Reference Monitor, Windows kernel, Intel, attacks on Defender, kernel data protection.

## 1. INTRODUCTION

Microsoft Windows is the dominating desktop operating system (OS) worldwide, and it is still facing a huge number of kernel-mode threats. The key challenge is that all loaded kernel drivers share the same memory space with the rest of the Windows OS kernel. There is no built-in solution to control and restrict all memory access between these drivers and sensitive kernel data. As a result, kernel driver-based attacks are still dangerous for Windows OS. The global trend of kernel threats is to bypass AV/EDR solutions by disabling or blinding them to achieve a permanent and undetectable malware presence on a computer.

Windows experts are well familiar with this challenge and continue developing various outstanding security solutions to restrict the scope of kernel-mode attacks. One of the integrated security mechanisms is called Driver Signature Enforcement (DSE), which is designed to restrict attackers from loading malware drivers.

DSE requires a special Kernel Mode Code Signing (KMCS) that allows loading kernel drivers with special digital code signatures only, and it is no longer possible to load unsigned drivers. Attackers can disable DSE by modifying kernel variable nt!g_CiEnabled (CI!g_CiOptions). Since Windows 8.1, this variable is protected by PatchGuard (Poslušný, 2022) and in Windows 11 this variable is protected by MmProtectDriverSection through static KDP (Hollow, 2021). Attackers can leverage the following techniques to gain kernel privileges:

- Load and exploit signed vulnerable drivers;
- Sign malware drivers by stolen digital certificates.
- Pass WHQL tests as legal drivers;

**Malware Abuse Signed Buggy Drivers.** According to Foster (2021) from CrowdStrike, "in a typical kernel attack, adversaries install and load a known vulnerable driver to gain access to the system". Two security points can be underlined. The first is that malicious actors can "leverage this vulnerable driver to access the kernel, where they can perform any number of actions, including deleting or deactivating security measures". The second point is that "since these drivers are legitimately signed and certificate revocation is difficult, unpatched versions of the drivers can be used "in the wild" for years without being detected or blocked".

Atsiv utility by Linchpin Labs is one of the first toolkits that allow loading of unsigned drivers in Windows Vista x64. Atsiv used its own PE loader that was not visible in Windows standard drivers list. On the one hand, it is a rootkit-type behavior and, on the other hand, the Atsiv driver is not malicious by itself. Microsoft blocked this tool by revoking its certificate (Espiner, 2007).

According to the MITRE (2021), the attack technique that abuses digitally signed vulnerable kernel drivers is called Bring Your Own Vulnerable Driver (BYOVD). Adversaries may include the vulnerable driver with files delivered to the target machine.

Baines (2021) from Rapid7 highlighted that BYOVD is a common technique used by advanced adversaries and opportunistic attackers alike. The author provided about 30 malware examples that use various legitimate drivers to prove that BYOVD is a valuable technique.

According to Poslušný (2022), ESET experts analyzed recent malware cases that utilized vulnerable drivers and note that this problem is not new, having been widely discussed in the past. Nevertheless, it is still a security issue as of this date. Here are some high-profile APT that used the BYOVD technique:

- Slingshot APT leverages the following buggy drivers Goad, SpeedFan, Sandra, and ElbyCDIO;
- InvisiMole APT leverages SpeedFan.sys;
- RobbinHood ransomware uses a Gigabyte motherboard driver.
- Moriya rootkit loads VirtualBox driver (Lechtik and Dedola, 2021).
- A data wiper malware named HermeticWiper abuses epmntdrv.sys, a signed driver of EaseUS Partition Master (Guerrero-Saade, 2022).

One more example of the BYOVD technique is KDMapper, a recently issued tool by Cruz (2021) that exploits the Intel driver (iqvw64e.sys) to manually map non-signed drivers in memory.

Hfiref0x issued several security tools: Turla Driver Loader named TDL (Hfiref0x, 2019) is designed to bypass Windows x64 Driver Signature Enforcement (DSE) and Kernel Driver Utility named KDU (Hfiref0x, 2022a) that uses vulnerable drivers as "providers" to disable DSE, map any driver, and execute it.

Pham et al. (2022) concerned that "many collections of vulnerable drivers are easily found on the Internet". Eclypsium researchers, Jesse and Shkatov (2019) found more than 40 vulnerable drivers, all them "come from trusted third-party vendors, signed by valid Certificate Authorities, and certified by Microsoft".

The security problems of vulnerable drivers can greatly impact the target system. SentinelLabs experts discovered five high severity bugs in the Dell driver from its BIOS update utility. Hackers can exploit these bugs to gain kernel-mode privileges to attack hundreds of millions of computers worldwide, including desktops, laptops, and tablets. The list of affected computers has over 380 models. One more key challenge is that these high severity vulnerabilities have remained undisclosed for more than 10 years (Dekel, 2021; Clark, 2021).

**Malware Drivers Signed by Stolen Certificates**. Usage of stolen, leaked or misused third-party certificates is a common attacker's technique. For example, DirtyMoe malware uses a driver signed with a revoked certificate. Chlumecký (2021) discovered that Windows allows loading drivers signed with revoked certificates even if an appropriate certificate revocation list (CRL) is locally stored.

According to the research by Barysevich (2018), Kozák et al. (2018), and Shoeb (2021) the Dark Web is one of the main platforms for selling code-signing certificates, including Extended Validation (EV) certificates to sign critical code such as Windows drivers.

Abrams (2022a) has found that recently leaked Nvidia code code-signing certificates have already been used by attackers to sign malware drivers.

Fortinet Sde-Or and Voronovitch (2022), researchers with Fortinet's FortiGuard Labs, analyzed novel kernel rootkit named 'Fire Chili', that is signed using stolen digital certificates from Frostburn Studios (game developer) or from Comodo (security software). The rootkit is designed to hide and protect malicious artifacts: files, processes, registry keys and network connections from user-space security solutions.

Kwon et al. (2021) investigated the characteristics of code-signing abuse and observed that this issue is quite prevalent in South Korea. They underlined that only 6.8% of the abused certificates are revoked, which leads to extending the validity of certified malware in the wild.

**Kernel-level Anti-Cheat Software**. One more kernel threat is coming from the game protection industry. Game developers issue a special system of software against cheats to keep the game fair and square. Modern anti-cheat engines include kernel-mode drivers, such as Ricochet Anti-Cheat protects Call of Duty and Easy Anti-Cheat protects more than 80 games and is installed by over 100 million PC players globally. Latić (2021) and Menegus (2022) are raising concern that bugs in such drivers can have a high impact on many users. For example, Genshin Impact, an action game, installs an anti-cheat driver named mhyprot2 to protect the game process. However, this driver can be used to get illegal read and write access to the kernel memory Oki (2021a).

Lechtik et al. (2021), security experts from Kaspersky Lab, revealed that the Cheat Engine driver (dbk64.sys) was used by Demodex rootkit from GhostEmperor's infection chain to bypass the Windows Driver Signature Enforcement mechanism.

**UEFI Security Threats**. Another big trend is UEFI firmware-level malware implants, which are usually highly targeted. MoonBounce revealed by Lechtik et al. (2022) from Kaspersky Lab is one of the recent stealthy UEFI rootkits that can inject a malicious driver into the Windows kernel. Security researchers from Binarly have found more than two dozen UEFI vulnerabilities that impact millions of devices (Kovacs, 2022).

**WHQL Scandal.** Since 2016, Microsoft has required all third-party drivers submitted via its Windows Hardware Quality Labs (WHQL) testing process to be digitally signed by Microsoft itself. WHQL opens the door to get the driver distributed through Windows Update or the Microsoft Update Catalog. Microsoft Hardware Certification is a rigorous process of drivers validation. However, according to Fingas (2021) and Vijayan (2021), Microsoft has accidentally signed several malware drivers in recent months.

Netfilter driver was the first rootkit discovered to be using WHQL digital signature issued by Microsoft directly. Despite connecting to malware C&C servers, this driver successfully passed through the Windows Hardware Compatibility Program (WHCP). MSRC (2021) has confirmed that the malicious driver has been accidentally signed.

FiveSys is one more malware driver with a Microsoft's valid WHQL digital signature. Bitdefender's experts reported that this malware is similar to the Netfilter and targeted online video games in China region for over a year (Istrate et al., 2021).

**Microsoft Vulnerable and Malicious Driver Reporting Center**. Microsoft (2021) launches a new Vulnerable and Malicious Driver Reporting Center to fight the high rise of attacks based on vulnerable signed drivers. Using Windows Defender Security Intelligence (WDSI), formerly known as Microsoft Malware Protection Center (MMPC) users can submit a driver that will be analyzed by the Microsoft automated scanner.

Using signed buggy drivers attackers can accomplish a wide variety of tasks, for example, they can access virtual kernel memory for reading and writing. Various worldwide researches such as Rui (2020), Malvica (2020), Stein (2020), Bs (2020), and VL (2020) showed that attackers can tamper with kernel callback routines to prevent notifying the AV/EDR solutions of things such as process creation, thread creation, image loading, which is crucial for endpoint security.

The security experts from Positive Technologies (2021) analyzed the recent trends of kernel-mode threats and highlighted that 77% of rootkits are used for espionage purposes. They concluded that the main task of modern kernel-mode rootkits is to prevent the detection of further malicious activity.

Pham et al. (2022) raised the security problem with vulnerable kernel-mode drivers. They mentioned research projects developed by _xeroxz to map unsigned code into the kernel memory using physical memory read and write permissions as well as highlighting the problem of using vulnerable drivers to bypass anti-cheat mechanisms. They underline the fact that attackers can abuse vulnerable drivers to bypass or blind security products without terminating them, which prevents revealing the fact of the attack.

Currently, we can conclude that the global kernel malware trend is to disable or blind security solutions. As the default Windows AV, Microsoft Defender is facing a huge rise in attacks.

**Threat model**. Let us assume that using various approaches, intruders can execute malicious kernel code to disable, blind, or terminate Microsoft Defender by reading and overwriting kernel data without triggering any security features, see Figure 1. We assume that attackers are not able to disable PatchGuard.

**Microsoft Windows Improving Defenses Against Growing Kernel Threats.** Microsoft continues expanding its kernel-mode protection by issuing Kernel Patch Protection (KPP), informally known as PatchGuard, and Secure Kernel Patch Guard (SKPG) also known as HyperGuard (HG) to restrict access attempts from kernel drivers to the sensitive kernel memory (Shafir, 2022). However, there are several open-source research projects designed to disable the PatchGuard:

- NoPatchGuardCallback by Oki (2021b);
- UPGDSED by hfiref0x (2019);
- EfiGuard by Lavrijsen (2021);
- Shark by Blindtiger (2021).

Windows 10 introduced a new security concept named Virtualization-based security, or VBS, which leverages Hyper-V, a Windows hypervisor, and hardware virtualization features to protect data using containers.

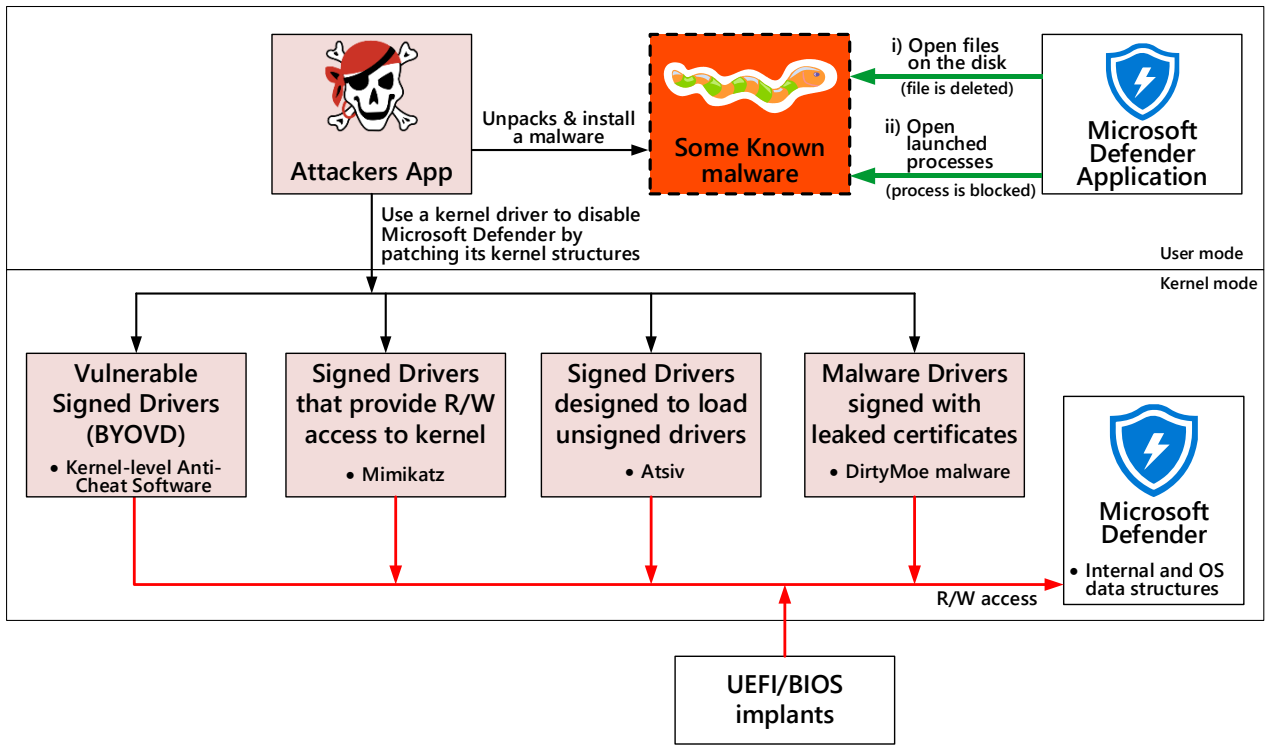

Figure 1. Attackers use the facilities of kernel-mode drivers to tamper with Microsoft Defender

Windows includes several new VBS features, such as Device Guard (DG) and Credential Guard (CG) that enhance the safety of the OS and user data using Hyper-V (Yosifovich et al., 2017).

VBS creates the Virtual Secure Mode (VSM), also known as Core Isolation, to segregate the most sensitive Windows services and data from attackers by using two Virtual Trust Levels (VTLs): VTL0 and VTL1.

VSM Secure Mode (VTL1) contains critical parts of the OS: Secure Kernel (SK) and Isolated User Mode (IUM) processes called trustlets. While VSM Normal Mode (VTL0) contains the main part of the OS kernel, all the rest of the drivers and applications including attackers' ones. VBS guarantees the security boundary between VTL0 and VTL1. VBS/Secure Kernel is currently renamed Windows Defender System Guard Container (Bisson, 2019).

Without enabled VBS, attackers can easily extract users' credentials from LSASS memory by disabling SRM and PPL mechanisms (Korkin, 2021). CG helps to prevent such attacks on users' credentials. CG allows storing credentials in memory of the Isolated Local Security Authentication Server (Lsaiso.exe), which is located in VTL 1 Trustlet.

CG helps to prevent one more important attack presented by Chilikov and Khorunzhenko (2015) and results in retrieving the cryptographic keys by accessing CNG!RandomSalt and CNG!g_ShaHash. Thanks to the CG, the kernel drivers are not able to access CNG variables, because these values are protected by SK (ERNW, 2019).

Device Guard (DG) is designed to protect machines from different kinds of software- and hardware-based attacks. DG leverages code integrity and enforces code integrity policies.

Kernel Data Protection (KDP) is one more mechanism designed to prevent attacks on kernel data. According to the Windows Base Kernel Team, KDP provides the ability to mark some kernel memory as read-only, preventing attackers from ever modifying protected memory (Allievi, 2020).

However, these hypervisor-based protection technologies include several limitations:

- VBS is mainly designed to protect Windows components. VBS does not provide APIs to isolate the memory of third-party drivers and apps;
- VBS supports only two isolated enclaves, without providing API functions to allocate more enclaves and provide fine-grained protection;
- KDP does not provide any API for developers to protect the memory of third-party drivers.

**HVCI Driver Blacklist and Microsoft Vulnerable Driver Blocklist**. Hypervisor-Protected Code Integrity (HVCI) is a security feature introduced in Windows 10. Enabled HVCI provides blacklist-based validation for each attempt to load a driver. Hfiref0x (2022b) analyzed this feature and concluded that this list does not include some known vulnerable drivers and should be bigger twice or triple. Microsoft VP D. Weston (2022) highlighted that Microsoft Defender for Windows 11 and 10 gains new security feature named Microsoft Vulnerable Driver Blocklist to address vulnerable drivers. This feature is based on virtualization-based security (Carnevale, 2022).

**Microsoft Defender is Under Attack Itself.** A serious VBS drawback is the lack of protection of critical processes, such as MsMpEng or the Antimalware Service Executable, which is an important part of Windows Security, formerly known as Windows Defender. This leads to the high rise of the various attacks on Microsoft Defender to disable and blind it.

Microsoft Defender is the primary AV on more than half a billion devices. Lefferts (2021), Corporate Vice President, Microsoft 365 Security, said that, according to Gartner, Microsoft is the Leader in the Endpoint Protection Platforms (EPP) Magic Quadrant.

T. Ganacharya, Partner Director for Security Research at Microsoft Defender for Endpoint, underlined that "Windows Defender is protecting more than 50% of the Windows ecosystem, so we're a big target, and everyone wants to evade us to get the maximum number of victims" (Tung, 2019).

Various security experts analyzed the internals of popular EDR/AV solutions, including Microsoft Defender, to bypass them (Karantzas and Patsakis 2021, Botacin et al.,2022).

The presented paper shows a new attack on Microsoft Defender based on abusing Windows security mechanisms by kernel driver without triggering PatchGuard.

The customized MemoryRanger will be used to prevent such attacks by restricting illegal access to the kernel data structure.

The remainder of the paper proceeds as follows.

Section 2 provides the analysis of existing attacks on Microsoft Defender that can result in disabling, terminating, and blinding Windows AV.

Section 3 presents the details of the proposed kernel attack on Microsoft Defender. This section includes an explanation of the involved Windows Internals mechanisms.

Section 4 contains the details of MemoryRanger customization to prevent these attacks.

Section 5 focuses on the main conclusions.

## 2. BACKGROUND ANALYSIS OF ATTACKS ON MICROSOFT DEFENDER

This section covers the analysis of various attacks on Microsoft Defender using opportunities of code running in user- and kernel- modes (Figure 2).

Microsoft Defender is a security software deeply integrated into the system with a lot of various parts. However, it is possible to separate user-mode modules, such as MsMpEng.exe, NisSrv.exe, MsMpEngCP.exe, and MpEngine.dll, and kernel-mode components, such as WdBoot.sys, WdDevFlt.sys, WdFilter.sys, WdNisDrv.sys. Some of them are deeply analyzed, such as MpEngine.dll by Bulazel (2018), WdFilter by Narib (2020).

Security researchers analyzed various components of Microsoft Defender or Microsoft Protection Service. Bulazel (2018) at the BlackHat USA presented the details of Microsoft Malware Protection Engine (mpengine.dll), which is one of the key user-mode components of Microsoft Defender. Experts from CyberArk investigated a file scanning process of Microsoft Defender (Dekel, 2017). Grabber (2018) analyzed the "Dynamic Code Security" mitigation of the Microsoft Defender Application.

Narib (2020) presented a series of posts about the internals of WdFilter.sys driver, which is the main kernel component of Microsoft Defender and provides File System Minifilter, and handlers operations in callbacks, such as process and thread creation; image loading, desktop handle and registry operations. Narib (2019) also analyzed the WdBoot.sys, a Microsoft Defender ELAM driver.

Vella (2019) at the CrikeyCon 2019 presented his results about reversing EDR solutions and analyzed their weaknesses.

Various research results regarding bypassing and evading Microsoft Defender were given by IredTeam (2019), purpl3f0x (2021).

**Malicious Software vs Microsoft Defender**. Malware is continuing to attack Microsoft Defender. The following malware examples will be reviewed: TrickBot Trojan, Zloader Trojan, DeroHE ransomware.

One of the recent versions of TrickBot Trojan uses a set of PowerShell commands to disable Microsoft Defender: deletes its service; terminates its process; abuses the Defender restriction policies; disables notifications and real-time protection (Maude, 2021; Abrams , 2022a). After that, TrickBot creates a scheduled task at the system startup to ensure persistence (MS-ISAC, 2020).

Beaume (2021) shares the PowerShell script that disables Microsoft Defender, including scanning engines via Set-MpPreference commands.

According to Cocomazzi and Pirozzi (2021), Zloader Trojan disables Microsoft Defender and its security modules such as Potentially Unwanted Applications (PUA) protection and Real-Time Monitoring, and adds exclusions, to hide all the malware components from Microsoft Defender:

- cmd /c powershell.exe -command "Set-MpPreference -MAPSReporting 0" to disable Microsoft Active Protection Service.
- powershell.exe -command "Add-MpPreference -ExclusionProcess "*.exe""

Lakshmanan (2021) highlighted that MosaicLoader malware uses similar PowerShell commands to exclude its installed folders from Defender inspections. The reverse analysis of MosaicLoader was done by Szeles (2021) from Bitdefender. The logic of excluding the EXE file from being scanned was analyzed by Gemzicki (2021).

Abrams (2021) analyzed DeroHE ransomware and concluded that it also hides itself from Microsoft Defender by adding its folder to the exclusion by using Windows Management Instrumentation Command-line (WMIC) tool, named WMIC.EXE:

- @WMIC /Namespace:\\root\Microsoft\Windows\Defender class MSFT_MpPreference call Add ExclusionPath=\"\Temp\\".

Bichet (2020), a security researcher from Intrinsec analyzed Egregor and Prolock ransomware. Egregor evades protection by creating a Group Policy Object to disable Microsoft Defender, which sets the flag DisableAntiSpyware. This flag is designed to disable Microsoft Defender Antivirus during the deployment of another antivirus product. However, due to the Tamper Protection feature, this flag is unavailable in the newest Windows OSes.

Sordum (2021) team issued a tool named Defender Control that disables Microsoft Defender permanently by setting the registry values, modifying the group policies, and stopping windefend service. The authors explained that their work is designed to free precious resources if another anti-malware protection system is installed.

The rest of the section is focused on recent research papers and blogs dealing with attacks on Microsoft Defender. All attacks can be divided into two groups: attacks in the kernel- and user- modes.

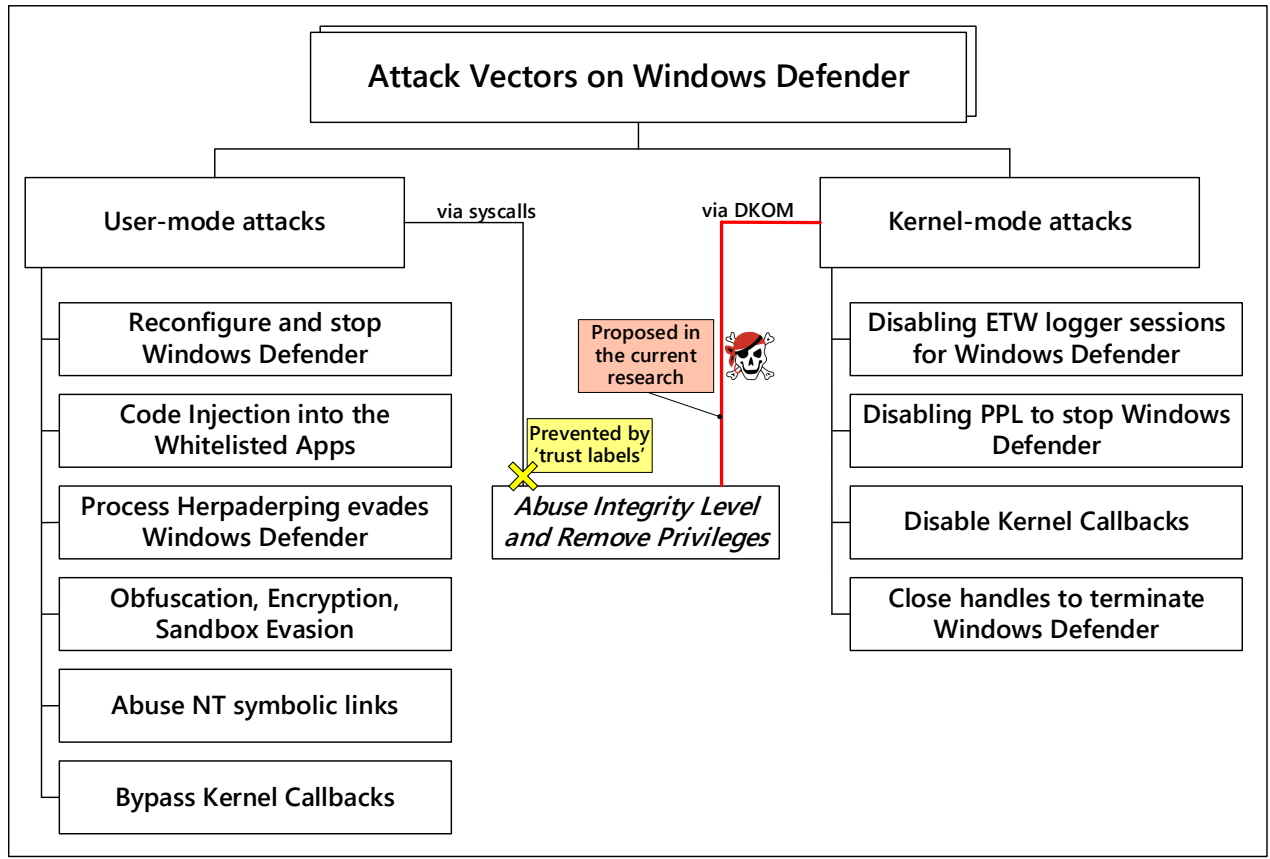

Figure 2. Attack Vectors on Microsoft Defender

### 2.1. KERNEL-MODE ATTACKS ON MICROSOFT DEFENDER

All kernel-mode attacks can be divided into the following subgroups:

- Attacks on ETW.
- Clear PPL to terminate the process.
- Disable and Bypass Kernel Callbacks.
- Terminate the Microsoft Defender Process by closing its handles.

**Attacks on ETW.** Microsoft Defender uses various Windows OS features: user-mode, kernel-mode components, and it leverages Event Tracing for Windows (ETW) facilities. ETW is a feature deeply integrated into the OS kernel. ETW was originally designed for performance troubleshooting, and it is currently used by various EDR solutions as a supplier of various OS-related events. ETW is widely used by common EDR/AV solutions. Microsoft Defender also uses ETW and gathers data from two ETW sessions, named DefenderApiLogger and DefenderAuditLogger. These two sessions are protected by Secure ETW (PPL mechanism) and cannot be stopped easily. Teodorescu, Korkin, and Golchikov (2021) from Binarly present a kernel-mode attack that results in disabling ETW sessions without triggering any reaction from PatchGuard. The problem is that ETW can be easily tampered with. Such attacks blind the whole class of EDR solutions, including Microsoft Defender.

**Clear PPL to Terminate the Process.** Microsoft Defender leverages the Protected Process Light (PPL) mechanism to protect its process from being terminated, as well as from code injections.

However, attackers can use a kernel-mode driver to clear the Process Protection level and after that, they can terminate the Microsoft Defender easily (Blaauwendraad et al., 2020). The authors admitted that "killing the process will notify the user of the machine that its antivirus program has been disabled and prompt the user to restart the service."

Thompson (2017) demonstrated how to terminate MsSence.exe, which is Microsoft Defender Advanced Threat Protection Service, by removing process protection via Mimikatz driver.

A similar attack that results in shutting down Microsoft Defender Antivirus was presented by Naceri (2021). However, apart from using kernel drivers, the author calls ChangeServiceConfig2W() to remove PsProtectSignerAntimalware-Light from the WinDefend service and after that unloads the WdFilter service.

One more driverless attack results in stopping Defender Service (Dosxuz, 2022). The author uses the Token Impersonation technique to escalate the privileges.

**Terminate the Microsoft Defender Process by closing its handles from the driver**. Yasser (2019) presented a tool named Backstab that can kill protected processes. The author leverages the Sysinternals Process Explorer (ProcExp) driver, which is signed by Microsoft and supports closing process handles, which leads to process termination. PPL mechanism guarantees that the protected processes cannot be terminated, even by apps running with admin privileges. However, using a driver such processes can be terminated, without triggering any security reaction. Backstab can terminate a Microsoft Defender that is running as a Protected Process with Protection level, which is equal to "PsProtectedSignerAntimalware-Light". One more feature is that attackers can duplicate the handle for the malware process.

**Disable Kernel Callbacks**. Microsoft Defender registered several kernel callbacks to be notified about various OS events. Examples of bypassing and removing kernel callbacks were given by Feichter (2020), Stein (2020), Forrer and Bauters (2021).

Karantzas and Patsakis (2021) specify the following techniques to disable kernel callbacks:

- Zeroing the address of the callback routine;
- Unregister the callback routine
- Patch the code of callback routine.

Bs (2020) presented a tool named CheekyBlinder, that leverages a signed vulnerable kernel-mode driver to find process callback functions and remove specific callbacks to blind EDR solutions. The author tested his tool using Avast Free AV, but the same techniques can be used to blind Microsoft Defender.

A similar attack on kernel callback routines was explained using the Mimikatz driver by Blaauwendraad et al., (2020) and Hand (2020). In addition, the authors proposed to overwrite the corresponding callback routine using RET instruction. However, KPP can trigger such code manipulations.

**Bypass Kernel Callbacks**. Using one of the user-mode code injection techniques, which is based on creating and mapping PE sections, attackers can spoof the image name received by a callback routine. As a result, the EDR receives a string with a fake legitimate EXE name, while the loaded malicious image will be hidden (Dylan, 2021).

## 2.2. USER-MODE ATTACKS ON MICROSOFT DEFENDER

**Reconfiguring Microsoft Defender**. Researchers continue the analysis of Microsoft Defender configuration, which can be used to disable it. Lyk (2020) has revealed that adding registry key and setting value results is stopping Defender from attaching to any volumes. WdFilter.sys can be unloaded by Process Hacker by using the list of loaded modules of the System:4 process.

**Code Injection into the Whitelisted Apps**. Microsoft Defender includes a built-in whitelist of some common process images, such as explorer.exe and smartscreen.exe. It is possible to implant a malware DLL into a whitelisted process and to perform malware actions using a basic code injection technique (T0mux, 2018).

**Microsoft Defender Evasion by Process Herpaderping**. Process Herpaderping is a technique released by Shaw (2021) to evade security products including Microsoft Defender. The author revealed that a callback registered via PsSetCreateProcessNotifyRoutine(Ex) "is invoked when the initial thread is inserted, not when the process object is created". Using this feature, attackers can confuse EDR/AV solutions by modifying on-disk file process content after the image has been mapped. The author successfully tested the proposed technique on a system with enabled Microsoft Defender on Windows 10.

**Obfuscation, Encryption, Sandbox Evasion**. Researchers reveal several ways to bypass Defender scanning using the following techniques:

- obfuscation (Defsecone 2020, Spinney 2019);
- payload encryption (Secarma, 2021)
- packing the payload (Unknow101, 2022);
- temporarily disabling the memory page access permissions (Billinis, 2020);
- sandbox evasion, payload encryption, and code injection (Born, 2021);

**Killing Defender by abusing NT symbolic links**. Security expert Lagrasta (2021), a member of Advanced Persistent Tortellini, revealed an interesting way to bypass Microsoft Defender. His idea is to apply NT symbolic links to temporarily redirect "\Device\BootDevice" to another disk, unload Windows Defender driver (WdFilter.sys), and, finally, load a fake driver from the substitute folder. The key point of this attack is that the researcher leverages Windows built-in mechanisms, pushing Microsoft Defender to follow the wrong path.

**Abuse Integrity Level and Remove Privileges**. One more attack based on Windows built-in mechanisms was presented by Landau (2022). An idea is to abuse Mandatory Integrity Control (MIC) and Security Reference Monitor (SRM) so that the Microsoft Defender become sandboxed from all the rest of the OS. The key security issue is that Windows provides the following syscalls to modify the protection features even for processes protected by PPL:

- SetTokenInformation() to minimize the Token Integrity Level.
- AdjustTokenPrivileges() to disable all the Token Privileges.

The author implemented a user-mode app named NerfToken.exe to reduce the Token Integrity level from system to untrusted and remove all privileges for Microsoft Defender. The corresponding proof-of-concept in C# was developed by TNP (2022) and in C developed by pwn1sher (2022).

Landau (2022) proposed a defense solution to block this attack by using another Windows built-in feature called "trust labels". The author explained that "trust labels allow Windows to restrict specific access rights to certain types of protected processes". The corresponding open-source proof-of-concept called PPLGuard designed by Elastic (2022) protects all running anti-malware PPL processes against this attack. PPLGuard is based on a userland exploit from PPLdump designed by Labro (2021) that abuse \KnownDlls. PPLGuard prevents disabling the Integrity Level by NerfToken-attack.

However, an attacker can use a kernel driver to implement a similar attack, that cannot be blocked by a user-mode defense solution, for example, PPLGuard. One more feature that can be improved is the number of removed privileges: Landau (2022) proposed to remove all privileges from Microsoft Defender, which is very suspicious and can be easily detected. An idea is to remove only important privileges, that help to disable Defender, without removing all privileges.

The next section covers the research results regarding the kernel attack based on Integrity Level and Remove Privileges, including the research results about Microsoft Defender Internals.

## 3. PROPOSED ATTACK ON MICROSOFT DEFENDER

This section describes the analysis of how Microsoft Defender detects malware before the attack on Microsoft Defender's kernel structures and after it.

### 3.1. RESEARCHING MICROSOFT DEFENDER INTERNALS

To research the internals of Microsoft Defender, a test bench was organized using the following steps:

- VMWare with installed Windows 11 x64 and WinDbg connected via COM port was used as a sandbox environment to research Microsoft Defender's internals during the malware detection.
- To trigger the reaction of Microsoft Defender and keep the detection under control, a zip password-protected archive with a malware file was used (Figure 3, *a*).
- The batch file was used to extract and launch the malware file (Figure 3, *c*).
- The Mimikatz app was used to trigger the malware detection engine of Microsoft Defender.
- To monitor the reaction of Microsoft Defender, the Process Monitor was used (Figure 3, *b*).

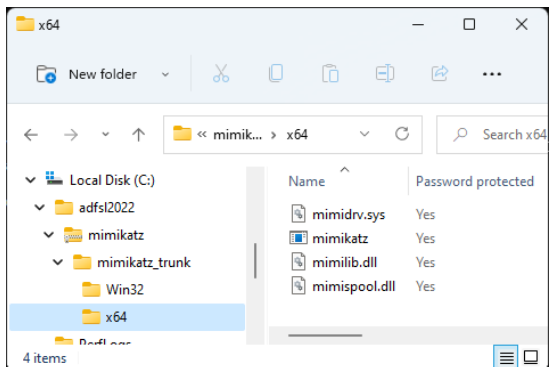
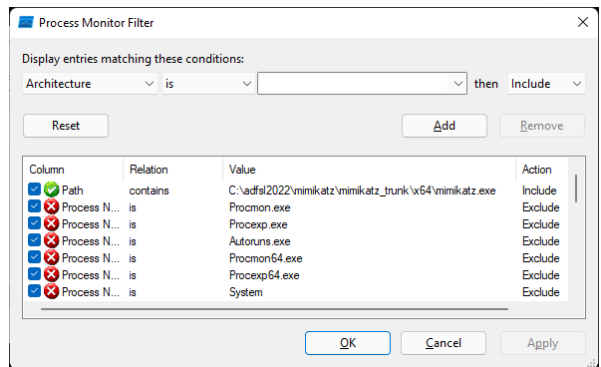

*a)*          *b)*

```
rmdir C:\adfsl2022\mimikatz /S/Q
7z.exe x C:\adfsl2022\mimikatz.zip -oC:\adfsl2022\mimikatz -pinfected
dir "C:\adfsl2022\mimikatz\mimikatz_trunk\x64"
"C:\adfsl2022\mimikatz\mimikatz_trunk\x64\mimikatz.exe"
```
*c)*

Figure 3. The testbench components: *a)* password-protected archive with malware (Mimikatz app)
*b)* Process Monitor with enabled Filter to monitor events only for the specified path
*c)* batch commands for extracting and launching the malware file from the specified path

### 3.2. DEFAULT BEHAVIOR OF MICROSOFT DEFENDER

First, the default behavior of Microsoft Defender was inspected using the aforementioned steps.

The fragments of ProcMon output show the key stages of experiments:

- Figure 4 shows that 7Z.EXE is extracting and writing the malware image file to the disk.
- Figure 5 shows that MsMpEng.EXE a core process of Microsoft Defender is reading the newly extracted malware file.
- Figure 6 ProcMon log shows that MsMpEng.exe is removing the detected malware file and after that, Microsoft Defender checks the file has been removed.

To remove malware file from the disk MsMpEng.EXE calls NtSetInformationFile() WinApi function, with input parameter FILE_DISPOSITION_INFORMATION_EX.Flags which is FILE_DISPOSITION_DELETE.

This flag indicates that the file is marked as deleted, and it will be deleted after the link count for the file became zero or all open handles for this file will be closed.

Microsoft Defender checks the file has been removed by calling CreateFile(). Figure 6 shows that CreateFile() function returns the status "DELETE_PENDING" and finally "NAME NOT FOUND", which indicates that the file is removed.

The next part covers the behavior of Microsoft Defender after the manipulation of its kernel structure.

figure 4. Step-1: ProcMon output shows that 7Z.EXE is extracting and writing the malware to the disk

figure 5. Step-2: ProcMon output shows that Microsoft Defender is reading the newly extracted file

figure 6. Step-3: ProcMon output shows that Microsoft Defender is removing the detected file and checking that the file has been removed

### 3.3. MANDATORY INTEGRITY CONTROL (MIC)

This section covers the internals of Mandatory Integrity Control (MIC) and how it can be used to restrict Microsoft Defender.

#### 3.3.1. Security Reference Monitor (SRM): Discretionary Access Control (DAC)

One of the main Windows security mechanisms is Security Reference Monitor (SRM), which implements Discretionary Access Control using access tokens and object security descriptors.

The information about process privileges is stored in the access token, which has a list of Security Identifier (SID). The security information about each object, for example, a file, a directory, or a registry key, is stored in the object security descriptor. An object security descriptor defines who can do what with this object.

When a program is accessing an object, SRM uses process token and object security descriptor to determine whether an access request should be granted. The security descriptor includes a Discretionary Access Control List (DACL), which comprises zero or more Access Control Entries (ACE). Each ACE includes SIDs and access rights for each SID.

Each time a process tries to get access to the object, for example, open a file or directory, an access request is generated. SRM processes an access request to grant or deny this access by comparing the SIDs from the access token and SIDs from the security descriptor located in DACL. This access control decision is implemented in the kernel-mode function nt!SeAccessCheck, which is one of the key functions of the Windows security model (Sandker, 2018). An example of an attack on token structures to bypass nt!SeAccessCheck was published in the JDFSL (Korkin, 2021).

Microsoft Defender as a system application executes with high privileges of the SYSTEM account with SID equals S-1-5-18 (NT Authority\System).

However, starting from Windows Vista, the SRM mechanism has been extended by adding a new feature named Mandatory Integrity Control (MIC). Microsoft Defender can be disabled by tampering with MIC. The next section covers the MIC details.

#### 3.3.2. Security Reference Monitor (SRM): Mandatory Integrity Control (MIC)

**MIC Overview**. Mandatory Integrity Control (MIC) expands DACL and is designed to isolate untrusted processes from the rest of the OS (Riley, 2006). MIC is defined by new integrity levels (ILs) that represent additional levels of trustworthiness and a mandatory policy to control access to objects. ILs are represented by Security Identifiers (SIDs).

Each object in Windows, such as a process, a registry key, or a file, has a separate IL, see Figure 7

Windows 11 supports the following ILs with the increasing privileges:

- Untrusted (0);
- Low (1);
- Medium (2);
- High (3);
- System (4);
- Protected (5).

Microsoft Defender is running with System IL. Core system apps, such as Services and Controller app, are running with High and System ILs, while Medium and High ILs are used by user's applications, such as Word, Skype, or Google Drive. To reduce the severity of web-based threats, web browsers (e.g. chrome) are running at low IL.

Each time any subject (e.g. process) attempts to interact with a target object (e.g. another process or a file), MIC checks the ILs of the initiator and the target.

MIC mandatory policy implements a Bell-LaPadula Model (BLP) and guarantees that the applications with low IL cannot get write or delete access to the objects with higher IL, even if the DACL-based SRM allows these access attempts. MIC restricts only reading process memory, it does not restrict read access to the files (Laiho, 2016).

A process can write to or delete an object only when its IL is equal or higher than the object's IL.

The next section covers the details of how MIC is implemented in Windows and how it can be used to disable Microsoft Defender.

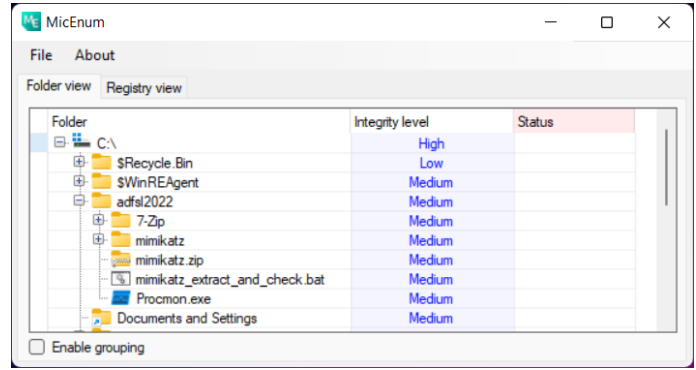

Figure 7. Newly created folders by default have an integrity level equals medium

**MIC Internals**. The research has revealed that Microsoft Defender periodically walks through the directories and opens processes and symbolic links. The corresponded WinAPI functions invoke nt!SeAccessCheck(), see Figure 8.

The pseudocode snippets of nt!SeAccessCheck() and nt!SepMandatoryIntegrityCheck() functions are presented in Figure 9 *a)* and *b)*.

For each running process, the information about its Integrity Level is stored in the field named IntegrityLevelIndex, which is located in the "EPROCESS.Token" structure.

Each time the process tries to access any object, the OS checks the privileges by calling nt!SeAccessCheck() and then nt!SepMandatoryIntegrityCheck(), that reads the IntegrityLevelIndex field.

The next paragraph shows an attack on Defender's Integrity Levels.

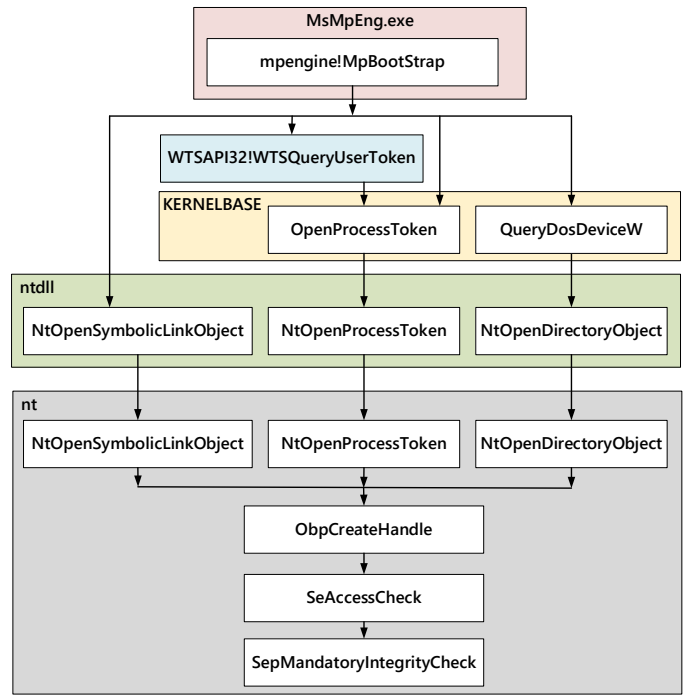

Figure 8. Defender calls functions that invoke SeAccessCheck() and SepMandatoryIntegrityCheck() during accessing file system

```
NTSTATUS SeAccessCheck(DesiredAccess,...)        NTSTATUS SepMandatoryIntegrityCheck(
{                                                   &AllowedAccessByMIC, &ProcessSIDfromMIC, ...)
    SepMandatoryIntegrityCheck(&AllowedAccessByMIC,...))   {
    if (DesiredAccess > AllowedAccessByMIC)          IntergrityLevelIndex = *(unsigned int *)(ProcessToken + 0xD0);
    {                                                if ( (_DWORD)IntergrityLevelIndex == -1 ||
        return = ACCESS_DENIED;                           (ppsid = (_WORD **)(*(_QWORD *)(ProcessToken + 0x98)
    }                                                     + 16 * IntergrityLevelIndex)) == 0i64 )
    else{                                            {
        if (SepAccessCheck()){                           ProcessSIDfromMIC = SeUntrustedMandatorySid;
            // access is granted                     }
        }else{                                       else
            // access is denied                      {
        }                                                ProcessSIDfromMIC = *ppsid;
    }                                                }
}                                                }
                    a)                                                    b)
```

Figure 9. The pseudocode snippets for the functions:
*a)* nt!SeAccessCheck() and *b)* nt!SepMandatoryIntegrityCheck()

**Patching IntergrityLevelIndex.** To restrict Microsoft Defender, we tried to overwrite the IntergrityLevelIndex field using (-1) or 0xFFFFFFFF value. This manipulation sets the lowest privilege (SeUntrustedMandatorySid) to the Defender process. Figure 10 *a)* and *b)* shows the Integrity Level of Microsoft Defender before and after patching.

After that we run the batch script to extract and run malware, see Figure 3, *a)* once more. Figure 11 shows the corresponding ProcMon output.

As a result, Microsoft Defender cannot overwrite or delete malware files. However, we still fail to launch an extracted malware file, the CreateFile API returns status 0xC0000906 (STATUS_VIRUS_INFECTED). This status means that "The operation did not complete successfully because the file contains a virus".

The research has revealed that this status is returned by WdFilter.sys, a Windows Defender driver.

WdFilter driver registers a mini-filter to monitor file operations via FltRegisterFilter. WdFilter prevents launching a malware file using Post-create callback by calling FltCancelFileOpen with status 0xC0000906 (STATUS_VIRUS_INFECTED) or 0xC0000022 (STATUS_ACCESS_DENIED).

Examples of mini-filters that protect files are implemented in HazardShield (xqrzd, 2015) and into the AvScan (Microsoft Corporation, 2015).

After Patching IntergrityLevelIndex Microsoft Defender did not show a Windows pop-up notification about the detected threat, however, the malware process still cannot be launched.

The reason for that is that Microsoft Defender still can get open access to the launched process to inspect its memory and blocked the process. The core anti-malware logic of Microsoft Defender is implemented in the user-mode library named mpengine.dll, running inside the MsMpEng process.

The next paragraph shows how to restrict Microsoft Defender from inspecting memory of another process.

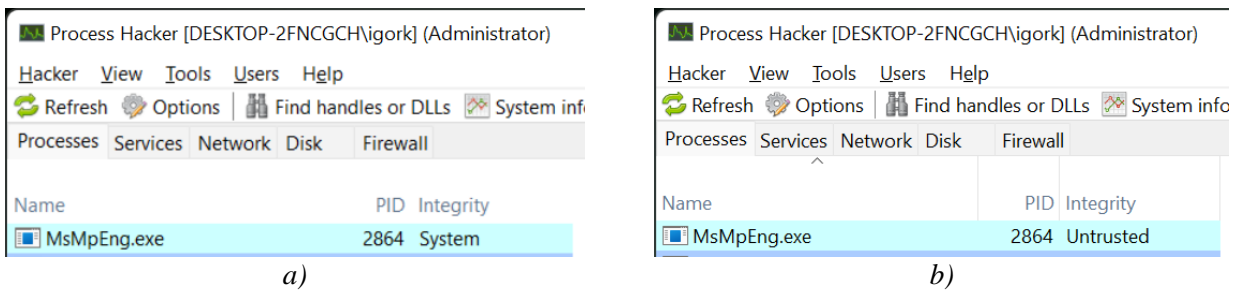

Figure 10. Integrity Level of Procmon: *a)* before patching IntergrityLevelIndex field and *b)* after it

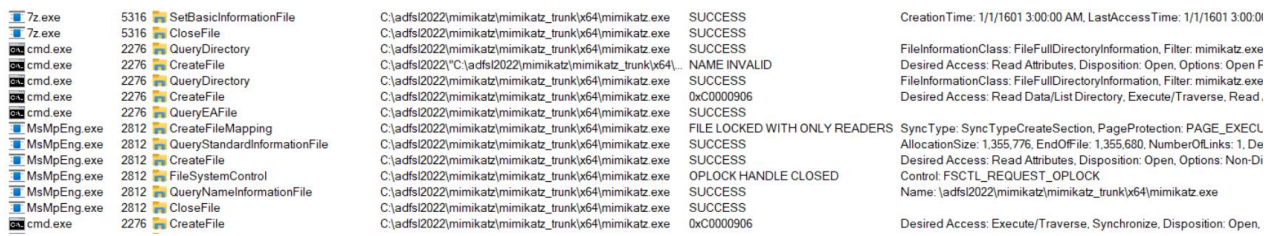

Figure 11. ProcMon output shows that cmd fails to open mimikatz file with error status 0xC0000906

### 3.4. TOKEN: REVOKE THE DEBUG PRIVILEGE

The research has revealed that Microsoft Defender inspects the newly launched process, the corresponding WinAPI functions are shown in Figure 12.

To restrict Microsoft Defender from inspecting a memory of newly created processes, attackers can revoke its Token Privileges. In the attack proposed by Landau (2022) all Microsoft Defender privileges were revoked. Our research task was to research and decrease the number of revoked privileges.

Information about process privileges is located in the kernel memory in the structure _SEP_TOKEN_PRIVILEGES, which is located in the Token from the EPROCESS structure. This structure includes three fields: Present, Enabled, and EnabledByDefault.

According to the MSDN (2021a), Chen (2008) "by default, users can debug only processes that they own. In order to debug processes owned by other users, you have to possess the SeDebugPrivilege privilege".

The research has revealed that clearing the SeDebugPrivilege bit from the field "Enabled", which is located in the "EPROCESS.Token.Privileges", is enough to prevent Microsoft Defender from inspecting the memory of newly launched malware processes and blocking them.

The next paragraph summarizes the attack steps and presents the tested results.

### 3.5. ATTACK'S SUMMARY AND TESTING

This research has revealed that Microsoft Defender detects and prevents malware propagation by using several stages:

I. At first, Microsoft Defender inspects the newly created files on the disk by opening the directory and files.
II. Second, Microsoft Defender inspects the newly launched process by inspecting their memory.

The proposed attack includes patching the following fields of the "EPROCESS.Token" structure corresponding to the Microsoft Defender:

I. IntergrityLevelIndex has to be overwritten by 0xFFFFFFFF or (-1).
II. Clear the SeDebugPrivilege bit from "Privileges. Enable".

The summary scheme of the attack is presented in Figure 13.

The experimental results show that these manipulations lead to extracting and launching malware using the script from Figure 3. The ProcMon output log shows that mimikatz has been successfully launched, see Figure 14.

Experiments show that attackers can execute even known malware by disabling Microsoft Defender without triggering any reaction from Windows security features, such as Kernel Patch Protector (KPP or PatchGuard).

The next section covers how to prevent this attack and protect Microsoft Defender.

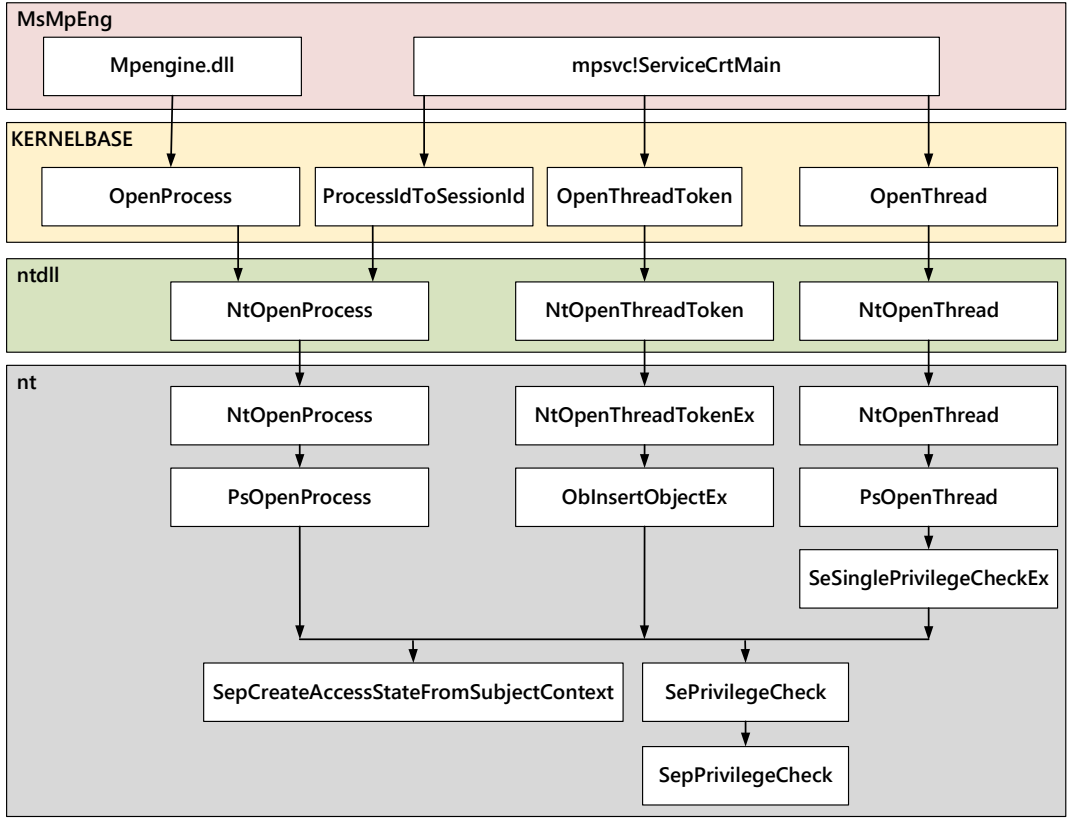

Figure 12. Defender calls functions that invoke SepPrivilegeCheck() and SepCreateAccessStateFromSubjectContext() during accessing launched processes

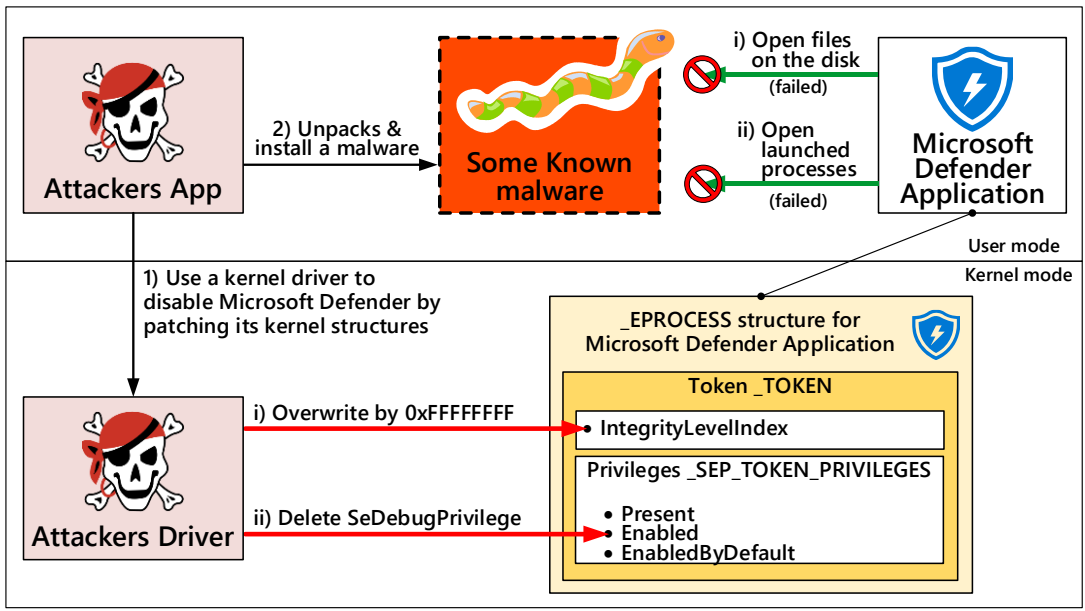

Figure 13. Malware app leverages a kernel driver to disable Microsoft Defender

| Process Name | PID | Operation | Path | Result | Detail |
|---|---|---|---|---|---|
| MsMpEng.exe | 2864 | ReadFile | C:\adfsl2022\mimikatz\mimikatz_trunk\x64\mimikatz.exe | SUCCESS | Offset: 196,608, Length: 65,536, Priority: Normal |
| MsMpEng.exe | 2864 | ReadFile | C:\adfsl2022\mimikatz\mimikatz_trunk\x64\mimikatz.exe | SUCCESS | Offset: 262,144, Length: 65,536, Priority: Normal |
| MsMpEng.exe | 2864 | ReadFile | C:\adfsl2022\mimikatz\mimikatz_trunk\x64\mimikatz.exe | SUCCESS | Offset: 327,680, Length: 65,536, Priority: Normal |
| MsMpEng.exe | 2864 | ReadFile | C:\adfsl2022\mimikatz\mimikatz_trunk\x64\mimikatz.exe | SUCCESS | Offset: 393,216, Length: 65,536, Priority: Normal |
| MsMpEng.exe | 2864 | ReadFile | C:\adfsl2022\mimikatz\mimikatz_trunk\x64\mimikatz.exe | SUCCESS | Offset: 458,752, Length: 65,536, Priority: Normal |
| MsMpEng.exe | 2864 | ReadFile | C:\adfsl2022\mimikatz\mimikatz_trunk\x64\mimikatz.exe | SUCCESS | Offset: 524,288, Length: 65,536, Priority: Normal |
| MsMpEng.exe | 2864 | ReadFile | C:\adfsl2022\mimikatz\mimikatz_trunk\x64\mimikatz.exe | SUCCESS | Offset: 589,824, Length: 65,536, Priority: Normal |
| MsMpEng.exe | 2864 | ReadFile | C:\adfsl2022\mimikatz\mimikatz_trunk\x64\mimikatz.exe | SUCCESS | Offset: 655,360, Length: 65,536, Priority: Normal |
| MsMpEng.exe | 2864 | ReadFile | C:\adfsl2022\mimikatz\mimikatz_trunk\x64\mimikatz.exe | SUCCESS | Offset: 720,896, Length: 65,536, Priority: Normal |
| MsMpEng.exe | 2864 | ReadFile | C:\adfsl2022\mimikatz\mimikatz_trunk\x64\mimikatz.exe | SUCCESS | Offset: 786,432, Length: 65,536, Priority: Normal |
| MsMpEng.exe | 2864 | ReadFile | C:\adfsl2022\mimikatz\mimikatz_trunk\x64\mimikatz.exe | SUCCESS | Offset: 851,968, Length: 65,536, Priority: Normal |
| MsMpEng.exe | 2864 | ReadFile | C:\adfsl2022\mimikatz\mimikatz_trunk\x64\mimikatz.exe | SUCCESS | Offset: 917,504, Length: 65,536, Priority: Normal |
| MsMpEng.exe | 2864 | ReadFile | C:\adfsl2022\mimikatz\mimikatz_trunk\x64\mimikatz.exe | SUCCESS | Offset: 983,040, Length: 65,536, Priority: Normal |
| MsMpEng.exe | 2864 | ReadFile | C:\adfsl2022\mimikatz\mimikatz_trunk\x64\mimikatz.exe | SUCCESS | Offset: 1,048,576, Length: 65,536, Priority: Normal |
| mimikatz.exe | 7684 | QueryNameInformationFile | C:\adfsl2022\mimikatz\mimikatz_trunk\x64\mimikatz.exe | SUCCESS | Name: \adfsl2022\mimikatz\mimikatz_trunk\x64\m |
| MsMpEng.exe | 2864 | ReadFile | C:\adfsl2022\mimikatz\mimikatz_trunk\x64\mimikatz.exe | SUCCESS | Offset: 1,114,112, Length: 65,536, Priority: Normal |
| MsMpEng.exe | 2864 | ReadFile | C:\adfsl2022\mimikatz\mimikatz_trunk\x64\mimikatz.exe | SUCCESS | Offset: 1,179,648, Length: 65,536, Priority: Normal |
| MsMpEng.exe | 2864 | ReadFile | C:\adfsl2022\mimikatz\mimikatz_trunk\x64\mimikatz.exe | SUCCESS | Offset: 1,245,184, Length: 65,536, Priority: Normal |
| MsMpEng.exe | 2864 | ReadFile | C:\adfsl2022\mimikatz\mimikatz_trunk\x64\mimikatz.exe | SUCCESS | Offset: 1,310,720, Length: 44,960, Priority: Normal |
| MsMpEng.exe | 2864 | CloseFile | C:\adfsl2022\mimikatz\mimikatz_trunk\x64\mimikatz.exe | SUCCESS | |
| MsMpEng.exe | 2864 | CloseFile | C:\adfsl2022\mimikatz\mimikatz_trunk\x64\mimikatz.exe | SUCCESS | |

Figure 14. ProcMon output shows that the mimikatz app has been successfully launched without terminating the Microsoft Defender process (MsMpEng.exe)

## 4. CUSTOMIZATION OF MEMORYRANGER TO PREVENT NEW ATTACKS

The presented attack on Microsoft Defender can be prevented by restricting illegal access to the kernel memory, which can be achieved by leveraging bare-metal hypervisors. Several open-source hypervisors can be used, such as Kernel-Bridge by HoShiMin (2021) or MemoryRanger by Korkin (2021). We decided to use MemoryRanger because it is designed to prevent attacks on kernel data. This section describes the details of how MemoryRanger can be customized to prevent disabling on Microsoft Defender.

**MemoryRanger Intro**. MemoryRanger (MR) is a software security solution designed to protect Windows OS kernel data from illegal access by kernel drives.

MR includes two main parts: a kernel-mode driver and a bare-metal hypervisor (type 1 hypervisor).

MR driver registers a few OS-level callback routines to be notified about various OS events: loading (unloading) drivers, creation (termination) processes. This driver sends information about really revealed sensitive data to the hypervisor.

A key part of MR is the bare-metal hypervisor that leverages Intel hardware-assisted virtualization technology (VT-x) and Extended Page Tables (EPT). One of the security features of the MR hypervisor is its ability to execute kernel drivers into isolated memory enclaves. Each enclave has a separate memory access configuration that restricts access to the kernel memory from drivers running inside the enclave.

MR has been chosen as a basic platform to protect kernel structures of Microsoft Defender.

**MemoryRanger Customization**. The following updates were added to the MemoryRanger to restrict access to its Token structure:

- MR driver locates EPROCESS structures for Microsoft Defender.
- MR driver reveals the addresses and sizes of the sensitive memory areas:
    - "Token.IntegrityLevelIndex" (4 bytes)
    - "Token.Privileges.Enabled" (8 bytes)
- MR driver traps loading of new drivers by installing a callback routine.
- MR hypervisor is notified about newly loaded drivers and creates a separate enclave for each of them. MR hypervisor restricts access to the sensitive memory areas for each loaded driver.

This scheme helps to trap loading of attacker's driver and prevents illegal access to the Microsoft Defender, see Figure 15. The scheme does not restrict access from OS core, such as ntoskrnl.exe, and all kernel drivers loaded before MR.

**MemoryRanger Testing and its Benchmark Assessment**. Experimental results show that this approach helps to prevent the attack with affordable performance degradation. The testbed has the following configuration:

- The computing testbed includes the host OS and VMware Workstation, which runs VM OS.
- PC with Intel i9-11900 CPU and 64 GB RAM is a host hardware platform.
- VM OS has been launched inside VMware using a CPU with 4 logical cores and 16 GB RAM.
- Windows 10, version 20H2 Build 10.0.19044 x64 is used for Host.
- Windows 11, version 20H2 Build 10.0.22000 x64 is used for VM.

## 5. CONCLUSIONS

To sum up, the following should be highlighted:

- Kernel-mode threats are still very dangerous for Windows OS. Attackers can exploit vulnerable drivers, sign malware drivers using leaked digital certificates. The existing Windows security features are not enough to restrict all these attack vectors.
- Global kernel malware trend is to bypass or blind security products without terminating the AV/EDR processes. As a result, Windows AV, Microsoft Defender is facing a huge rise in such attacks.
- Microsoft Defender can be attacked by several vectors. One of the recently published attacks abuses its Integrity level and removes all process privileges via WinAPI functions. However, this attack can be stopped by using trust labels.
- Nevertheless, by patching kernel data structures of Microsoft Defender attackers can disable it, without terminating any of its processes and without triggering any security features, e.g. PatchGuard.
- The customized MemoryRanger prevents this attack. Both an attack and its prevention has been tested on recent Windows 11.

**Preventing kernel-mode threats. Basics.** In modern operating systems such as Windows OS and Unix-based OS kernel-mode drivers share the same memory with the OS kernel core. This helps to improve the overall performance, but it also raise a security challenge with untrusted kernel components that can be used by attackers.

The general solution of preventing kernel attacks in modern OS is to isolate trusted kernel components from untrusted ones. Microsoft experts developed a Hyper-V that allow to run two separate kernel enclaves, named VTL0 (or Normal Mode) and VTL1 (or Secure Mode).

The Windows OS kernel core and all drivers are running into the VTL0. OS built-in kernel sensitive components such as Secure Kernel Code Integrity (skci.dll) and Kernel Mode Cryptographic Primitives Library (cng.sys) are loaded into VTL1.

This Microsoft scheme protects the memory of sensitive built-in components from illegal access by malware drivers. However, it does not provide isolation of third-party drivers and it support just two kernel enclaves

The designed MemoryRanger dynamically allocates a separate kernel enclave for each kernel driver and support flexible memory access restriction.

At the same time, there are several research solutions that leverage hypervisor-based technology to create isolated enclaves to isolate drivers:

- one enclave used in HACS by Wang et.al. (2017), and AllMemPro by Korkin (2018-a);
- two enclaves used in LKMG by Tian et.al. (2018), and EPTI by Hua et.el. (2018);
- three enclaves used in LAKEED by Tian et.al. (2017);

However, placing all OS drivers into separate isolated enclaves will cause a serious performance degradation.

The security research on blocking kernel-mode threats in modern OSes is still in progress.

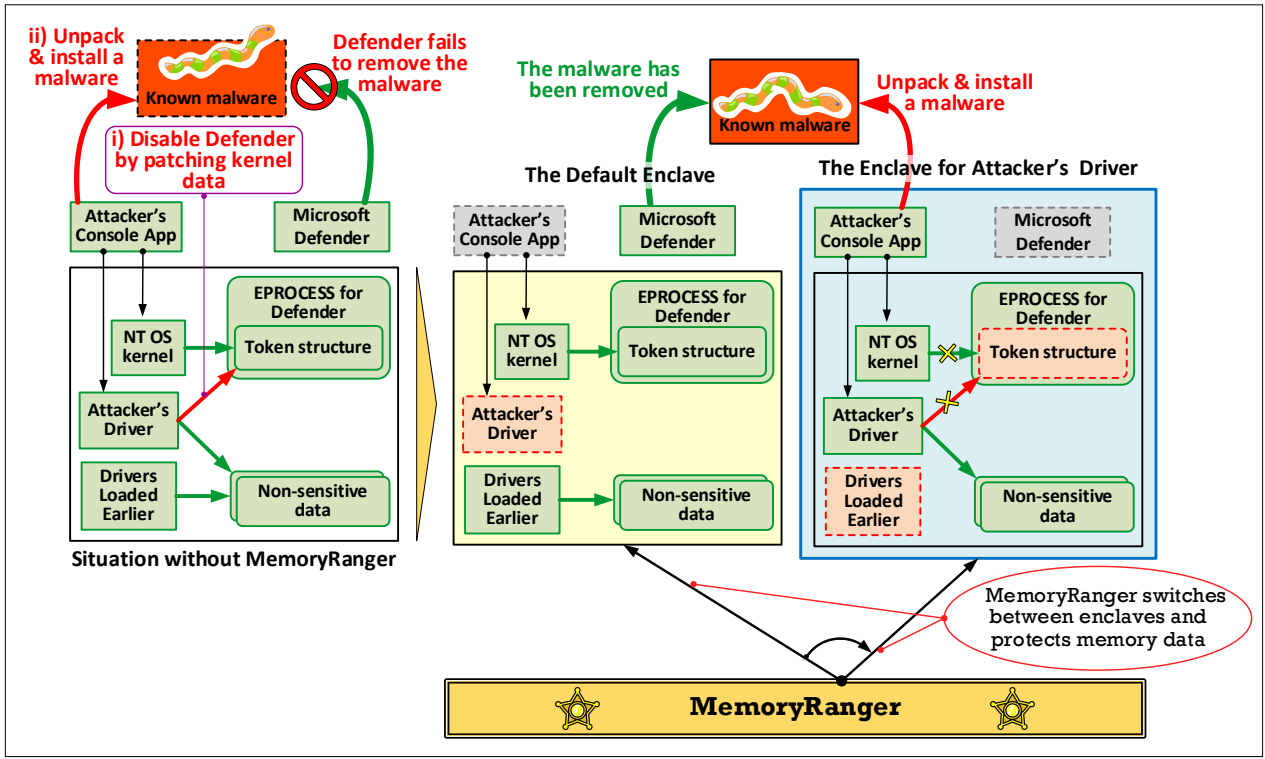

Figure 15. MemoryRanger protects the Token structure fields of the Microsoft Defender application from being patched by an attacker's driver using two enclaves: the default one for drivers loaded earlier and a separate enclave for newly loaded attacker's driver

## 6. APPENDIX A – TESTING ATTACK ON AV

The proposed attack was tested using some popular AV solutions, see Table 1. We can see that the proposed attack has blinded vast majority of AVs.

Table 1. Summary table of attacked AV solutions.

| AV Name | AV ability to detect malicious files | AV ability to detect malicious processes |
|---|---|---|
| Microsoft Defender | Disabled | Disabled |
| McAfee | Disabled | Disabled |
| Malwarebytes | Disabled | Disabled |
| Avast | Disabled | Disabled |
| AVG | Disabled | Disabled |
| Kaspersky | Disabled | Enabled |
| Trend Micro | Active, but AV cannot remove malware files | Disabled |

## 7. APPENDIX B – USING METASPLOIT TO TRIGGER MICROSOFT DEFENDER

A payload sample generated by the Metasploit Project can also be used to trigger Microsoft Defender and proof that it has been disabled.

The Metasploit Project is a computer security project that provides information about security vulnerabilities, IDS signature development, and aids in developing and using exploit code. This project includes several sub-projects: the Opcode Database, tools for evasion and anti-forensic, shellcode archive, and other research tools.

Metasploit was created by H. D. Moore in 2003 as a portable network tool using Perl. The project was released in 2004, it is completely free. By 2007, the Metasploit Framework had been completely rewritten in Ruby. In 2009 the project was acquired by Massachusetts-based security company Rapid7.

Metasploit is an open-source tool for developing and executing exploit code against a remote target machine. According to Kennedy et al. (2011), "this open-source platform provides a consistent, reliable library of constantly updated exploits and offers a complete development environment for building new tools and automating every aspect of a penetration test" (p. 16). The project includes about 600 payloads, which can run scripts or arbitrary commands against the host, grab the screen, upload and download files, evade antivirus defense, enable static IP address/port forwarding. The framework can be installed on macOS, Windows, and Linux.

Msfvenom is a command-line tool from the Metasploit package. The tool is used for generating various types of payloads. It provides the set of variable payloads from Metasploit, types of payload encodings, and various output file types.

According to Clarke (2020), "One example of using msfvenom in Kali Linux is to use it to create a malicious program that will connect the victim's system to your pentest system (a reverse shell), enabling you to obtain a meterpreter session with the target" (p. 139). Meterpreter session is a session with additional environment features, such as unified commands for all types of OS, the ability to upload and download files, including modules for post-exploitation.

For testing, we created a payload sample for Windows using the type "payload/windows/messagebox" without additional encodings. The sample shows a message box that contains the following text: "Hello, from MSF!". Here is the command for generating the payload for x86 Windows:

> "msfvenom -a x86 --platform windows -p windows/messagebox TEXT="Hello, from MSF!" -f exe > mes.exe"

According to Ortiz (2020), "Msfvenom creates payloads with common signatures that are picked up by almost all anti-virus solutions". As a result, without applying obfuscation techniques, the "malware codes are detected by several commercial antivirus packages. However, after applying different obfuscation methods, detection is much harder" (Palacios and Pérez-Sánchez, 2022). Microsoft Defender reveals the payload samples created without custom encryption and defines them as viruses. This example will be used to check whether or not Microsoft Defender is enabled.

Without tampering with Microsoft Defender, the generated payload sample is detected by virus and threat protection, see Figure 16.

The loaded 32-bit driver modifies the integrity level and downgrades the privileges of Microsoft Defender. The payload sample has not been blocked by Microsoft Defender, see Figure 17.

Microsoft Defender from Windows 10 32 bit has the same vulnerability and can be disabled by patching its EPROCESS.

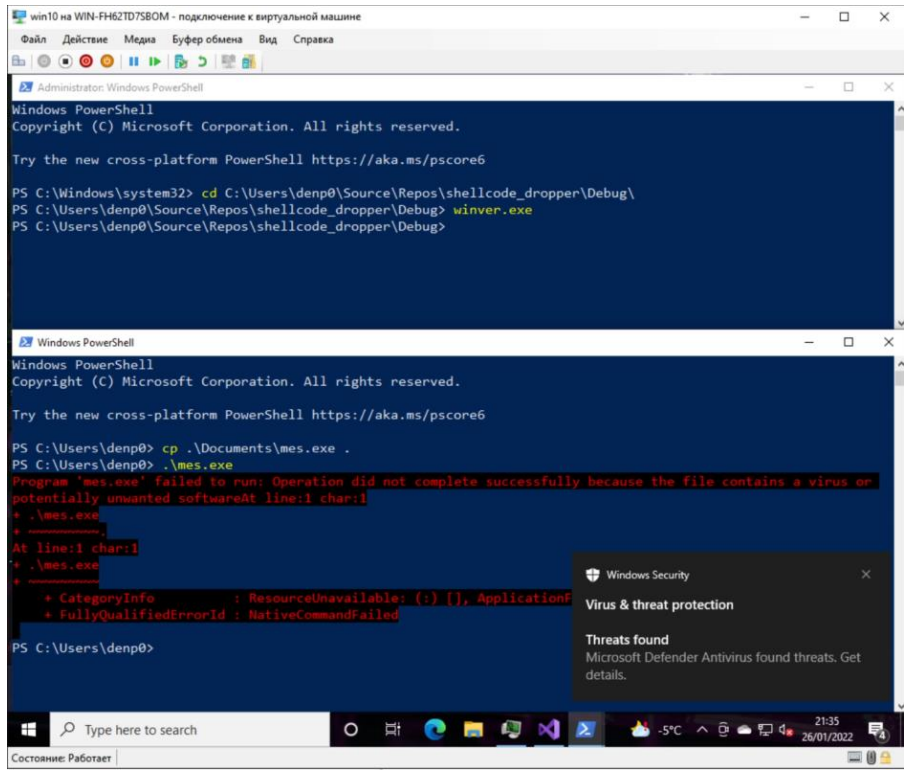

Figure 16. The payload has been detected and blocked by Microsoft Defender

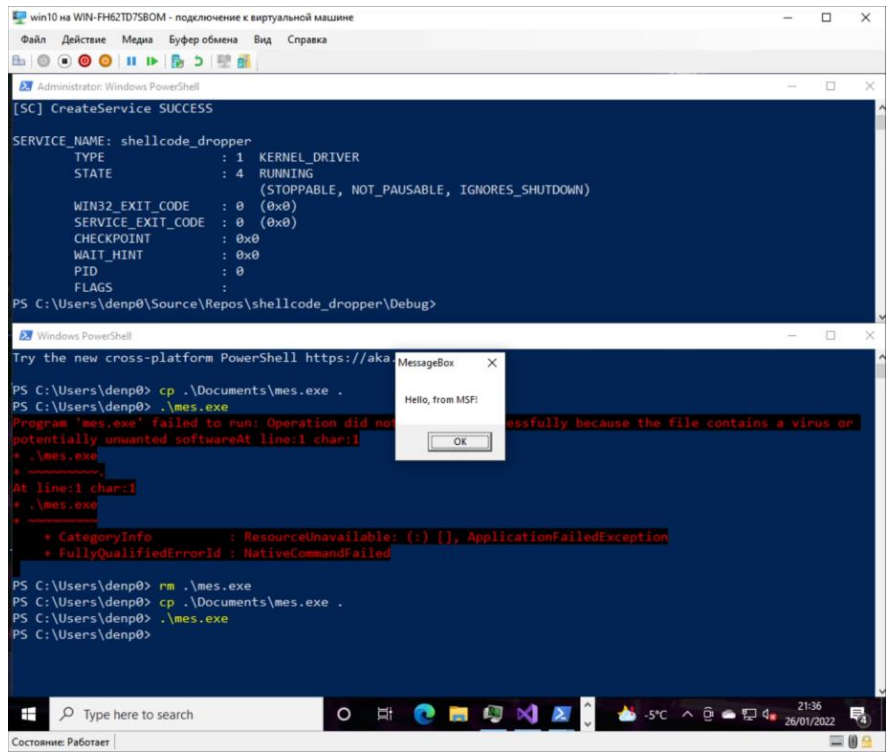

Figure 17. The payload has not been detected by Microsoft Defender.

## 8. ACKNOWLEDGMENTS

We thank the anonymous reviewers for their constructive feedback on this work.

## 9. REFERENCES

[1] Abrams, L. (2021). IObit forums hacked to spread ransomware to its members. BleepingComputer. Retrieved from https://www.bleepingcomputer.com/news/security/iobit-forums-hacked-to-spread-ransomware-to-its-members/

[2] Abrams, L. (2022a, March 5). Malware now using NVIDIA's stolen code signing certificates. BleepingComputer. Retrieved from https://www.bleepingcomputer.com/news/security/malware-now-using-nvidias-stolen-code-signing-certificates/

[3] Abrams, L. (2022b, July 30). New TrickBot Version Focuses on Microsoft's Windows Defender. BleepingComputer. Retrieved from https://www.bleepingcomputer.com/news/security/new-trickbot-version-focuses-on-microsofts-windows-defender/

[4] Allievi, A. (2020). Introducing Kernel Data Protection, a new platform security technology for preventing data corruption. Security Kernel Core Team. Retrieved from https://www.microsoft.com/security/blog/2020/07/08/introducing-kernel-data-protection-a-new-platform-security-technology-for-preventing-data-corruption/

[5] Baines, J. (2021). Driver-Based Attacks: Past and Present. Retrieved from https://www.rapid7.com/blog/post/2021/12/13/driver-based-attacks-past-and-present/

[6] Barysevich, A. (2018). The Use of Counterfeit Code Signing Certificates Is on the Rise. Recorded Future. Retrieved from https://www.recordedfuture.com/code-signing-certificates/

[7] Beaume, J. (2021). Disable Windows Defender in PowerShell – a script to finally get rid of it. Retrieved from https://bidouillesecurity.com/disable-windows-defender-in-powershell/

[8] Bichet, J. (2020). Egregor – Prolock: Fraternal Twins?. Retrieved from https://www.intrinsec.com/egregor-prolock/?cn-reloaded=1&cn-reloaded=1

[9] Billinis, C. (2020). Bypassing Windows Defender Runtime Scanning. Retrieved from https://labs.f-secure.com/blog/bypassing-windows-defender-runtime-scanning/

[10] Bisson, S. (2019). How virtualisation is changing Windows application security. Retrieved from https://www.techrepublic.com/article/how-virtualisation-is-changing-windows-application-security/

[11] Blaauwendraad, B., Ouddeken, T., Bockhaven, C. (2020). Using Mimikatz' driver, Mimidrv, to disable Windows Defender in Windows. Retrieved from https://rp.os3.nl/2019-2020/p61/report.pdf

[12] Blindtiger. (2021). Turn off PatchGuard in real-time for Win7 (7600) ~ later. Retrieved from https://github.com/9176324/Shark

[13] Born, G. (2021). Windows 11: Defender bypass with sandbox evasion. Retrieved from https://borncity.com/win/2021/10/19/windows-11-defender-bypass-mit-ausbruch-aus-der-sandbox/

[14] Botacin, M, Domingues, F. D., Ceschin, F., Machnicki, R., Antonio, M., Lício, P., Grégio, A. (2022). AntiViruses under the microscope: A hands-on perspective. Journal of Computers & Security. 112 (C), DOI: https://doi.org/10.1016/j.cose.2021.102500

[15] Bs. (2020). Removing Kernel Callbacks Using Signed Drivers. Retrieved from https://br-sn.github.io/Removing-Kernel-Callbacks-Using-Signed-Drivers/

[16] Bs. (2020). Removing Kernel Callbacks Using Signed Drivers. Retrieved from https://br-sn.github.io/Removing-Kernel-Callbacks-Using-Signed-Drivers/

[17] Bulazel, A. (2018). Windows Offender: Reverse Engineering Windows Defender's Antivirus Emulator. BlackHat USA. Retrieved from https://i.blackhat.com/us-18/Thu-August-9/us-18-Bulazel-Windows-Offender-Reverse-Engineering-Windows-Defenders-Antivirus-Emulator.pdf

[18] Carnevale. (2022). Microsoft Defender for Windows 11 and 10 gains new security feature that addresses vulnerable drivers. Windows Central. Retrieved from https://www.windowscentral.com/microsoft-defender-windows-11-and-10-gains-new-security-feature-addresses-vulnerable-drivers

[19] Chen, R. (2008). If you grant somebody SeDebugPrivilege, you gave away the farm. DevBlogs Microsoft. Retrieved from https://devblogs.microsoft.com/oldnewthing/20080314-00/?p=23113

[20] Chilikov, A., Khorunzhenko, E. (2015). Forensic analysis of RAM: detection, decryption and interpretation of encrypted cryptographic objects and data. RusCrypto 2015. Retrieved from https://www.ruscrypto.ru/resource/archive/rc2015/files/09_chilikov_khorunzhenko.pdf

[21] Chlumecký, M (2021). DirtyMoe: Rootkit Driver. Retrieved from https://decoded.avast.io/martinchlumecky/dirtymoe-rootkit-driver/

[22] Clark, M. (2021). Dell is issuing a security patch for hundreds of computer models going back to 2009. Retrieved from https://www.theverge.com/2021/5/4/22419474/dell-security-patch-kernel-level-permissions-firmware-update-driver-dbutil-sys

[23] Clarke, G. (2020). CompTIA PenTest+ Certification For Dummies. Wiley

[24] Cocomazzi, A., Pirozzi, A. (2021). Hide and Seek | New Zloader Infection Chain Comes with Improved Stealth And Evasion Mechanisms. SentinelLABS Research Team. Retrieved from https://assets.sentinelone.com/sentinellabs/SentinelLabs-Zloader

[25] Defsecone. (2020). Evading Windows Defender using obfuscation techniques. Retrieved from https://medium.com/@defsecone/evading-windows-defender-using-obfuscation-techniques-2494b2924807

[26] Dekel, K. (2017). Illusion Gap – Antivirus Bypass Part 1. Retrieved from https://www.cyberark.com/resources/threat-research-blog/illusion-gap-antivirus-bypass-part-1

[27] Dekel, K. (2021). CVE-2021-21551- Hundreds Of Millions Of Dell Computers At Risk Due to Multiple BIOS Driver Privilege Escalation Flaws. Retrieved from https://www.sentinelone.com/labs/cve-2021-21551-hundreds-of-millions-of-dell-computers-at-risk-due-to-multiple-bios-driver-privilege-escalation-flaws/

[28] Dosxuz. (2022). Stop Defender Service using C# via Token Impersonation. Retrieved from https://github.com/dosxuz/DefenderStop

[29] Dylan. (2021). Bypassing Image Load Kernel Callbacks. MDSec. Retrieved from https://www.mdsec.co.uk/2021/06/bypassing-image-load-kernel-callbacks/

[30] Elastic. (2022). PPLGuard is a proof of concept tool that can mitigate two currently-unpatched Windows security flaws. Retrieved from https://github.com/elastic/PPLGuard

[31] ERNW. (2019). Work Package 6: Virtual Secure Mode. SiSyPHuS Win10 Project. Retrieved from https://www.bsi.bund.de/SharedDocs/Downloads/DE/BSI/Cyber-Sicherheit/SiSyPHus/Workpackage6_Virtual_Secure_Mode.pdf?__blob=publicationFile&v=1

[32] Espiner, T., (2007). Microsoft blocks Vista driver 'hack' tool. ZDNET. Retrieved from https://www.zdnet.com/article/microsoft-blocks-vista-driver-hack-tool/

[33] Feichter, D. (2020). EPP/EDR Unhooking their protections. Retrieved from https://deepsec.net/docs/Slides/2020/EPP:EDR-Unhooking_Their_Protections_Daniel_Feichter.pdf

[34] Fingas, J. (2021). Microsoft signed a driver loaded with rootkit malware. Engadget Blog. Retrieved from https://www.engadget.com/microsoft-signed-netfilter-malware-driver-164228266.html

[35] Forrer, S. and Bauters, J. (2021). Kernel Karnage – Part 1. NVISO Labs. Retrieved from https://blog.nviso.eu/2021/10/21/kernel-karnage-part-1/

[36] Foster, B. (2021). Detecting and Preventing Kernel Attacks. Retrieved from https://www.crowdstrike.com/blog/how-to-detect-and-prevent-kernel-attacks-with-crowdstrike/

[37] Gemzicki, M. (2021). How I Evaded Windows Defender With A Batch Script — Kinda. Retrieved from https://devdotpy.medium.com/how-i-evaded-windows-defender-with-a-batch-script-kinda-fd21acae35d2

[38] Grabber, M. (2018). Documenting and Attacking a Windows Defender Application

Control Feature the Hard Way — A Case Study in Security Research Methodology. Retrieved from https://posts.specterops.io/documenting-and-attacking-a-windows-defender-application-control-feature-thehard-way-a-case-73dd1e11be3a

[39] Guerrero-Saade, J. A. (2022, February 23). HermeticWiper | New Destructive Malware Used in Cyber Attacks on Ukraine. SentinelOne. Retrieved from https://www.sentinelone.com/labs/hermetic-wiper-ukraine-under-attack/

[40] Hand, M. (2020). Mimidrv In-Depth: Exploring Mimikatz's Kernel Driver. Retrieved from https://posts.specterops.io/mimidrv-in-depth-4d273d19e148

[41] Hfiref0x (2019). Driver loader for bypassing Windows x64 Driver Signature Enforcement. Retrieved from https://github.com/hfiref0x/TDL

[42] Hfiref0x (2019). Universal PatchGuard and Driver Signature Enforcement Disable. Retrieved from https://github.com/hfiref0x/UPGDSED

[43] Hfiref0x (2022a). Kernel Driver Utility. Retrieved from https://github.com/hfiref0x/KDU

[44] Hfiref0x (2022b). KDU v1.2 release and the wonderful world of Microsoft incoherency. Retrieved from https://swapcontext.blogspot.com/2022/02/kdu-v12-release-and-wonderful-world-of.html

[45] Hollow. (2021). changes to ci.dll, g_CiOptions. Unknowncheats. Retrieved from https://www.unknowncheats.me/forum/general-programming-and-reversing/457661-changes-ci-dll-g_cioptions.html

[46] IredTeam. (2019). Evading Windows Defender with 1 Byte Change. Retrieved from https://www.ired.team/offensive-security/defense-evasion/evading-windows-defender-using-classic-c-shellcode-launcher-with-1-byte-change

[47] Istrate, C.A., Biro, B., Bleotu, R. C., Coblis, C. S. (2021). Digitally-Signed Rootkits are Back – A Look at FiveSys and Companions. Bitdefender. Retrieved from https://www.bitdefender.com/files/News/CaseStudies/study/405/Bitdefender-DT-Whitepaper-Fivesys-creat5699-en-EN.pdf

[48] Jesse, M., and Shkatov, M. (2019). Get Off the Kernel if You Can't Drive. DEF CON 27 Conference. Retrieved from https://eclypsium.com/2019/08/10/screwed-drivers-signed-sealed-delivered/

[49] Karantzas, G., Patsakis, C. (2021). An Empirical Assessment of Endpoint Detection and Response Systems against Advanced Persistent Threats Attack Vectors. Journal of Cybersecurity and Privacy, 1 (3), pp 387-421, Basel, Switzerland, DOI: https://doi.org/10.3390/jcp1030021

[50] Kennedy, D., Gorman, J., Kearns, D., and Aharoni, M. (2011). Metasploit: The Penetration Tester's Guide. No Starch Press; 1st edition

[51] King, D. (2016). Analysis of Windows Access Permission Checking Mechanism. ProgrammerSought. Retrieved from https://www.programmersought.com/article/3880644118/

[52] Korkin, I. (2021, June 10). Kernel Hijacking Is Not an Option: MemoryRanger Comes to The Rescue Again. Paper presented at the Journal of Digital Forensics, Security and Law: Vol 16, No.1, Article 4. Retrieved from https://commons.erau.edu/jdfsl/vol16/iss1/4

[53] Korkin, I. (2021, May 24-27). Protected Process Light is not Protected: MemoryRanger Fills the Gap Again. Paper presented at the Systematic Approaches to Digital Forensic Engineering (SADFE) International Workshop in conjunction with the 42nd IEEE Symposium on Security and Privacy. in Proceedings of 2021 IEEE Symposium on Security and Privacy Workshops, San Francisco, CA, USA, May 24-27, 2021, pp.298-308, Retrieved from https://conferences.computer.org/sp/pdfs/spw/2021/893400a298.pdf doi.org/10.1109/SPW53761.2021.00050

[54] Kovacs, E. (2022). Two Dozen UEFI Vulnerabilities Impact Millions of Devices From Major Vendors. SecurityWeek. Retrieved from https://www.securityweek.com/two-dozen-uefi-vulnerabilities-impact-millions-devices-major-vendors

[55] Kozák, K., Kwon, B., J., Kim, D., Gates, G., Dumitras, T. (2018). Issued for Abuse:

Measuring the Underground Trade in Code Signing Certificates. Retrieved from http://users.umiacs.umd.edu/~tdumitra/papers/WEIS-2018.pdf

[56] Kwon, B., Hong, S., Jeon, Y., Kim, D. (2021) Certified Malware in South Korea: A Localized Study of Breaches of Trust in Code-Signing PKI Ecosystem. International Conference on Information and Communications Security. DOI: http://dx.doi.org/10.1007/978-3-030-86890-1_4

[57] Labro, C. (2021). Dump the memory of a PPL with a userland exploit. Retrieved from https://github.com/itm4n/PPLdump

[58] Lagrasta, F. (2021). The dying knight in the shiny armour. Killing Defender through NT symbolic links redirection while keeping it unbothered. Retrieved from https://aptw.tf/2021/08/21/killing-defender.html

[59] Laiho, S. (2016). How Windows Security really works. Workplace Ninja User Group Switzerland. Retrieved from https://digiblog.s3-eu-central-1.amazonaws.com/app/20160203095837/How-Windows-Security-really-works-CMCE1602-Sami-Laiho.pdf

[60] Lakshmanan, R. (2021). This New Malware Hides Itself Among Windows Defender Exclusions to Evade Detection. The Hacker News. Retrieved from https://thehackernews.com/2021/07/this-new-malware-hides-itself-among.html

[61] Landau., G. (2022). Sandboxing Antimalware Products for Fun and Profit. Elastic Security Research. Retrieved from https://elastic.github.io/security-research/whitepapers/2022/02/02.sandboxing-antimalware-products-for-fun-and-profit/article/

[62] Latić, D. (2021). What kernel-level anti-cheat is and why you should care. LEVVVEL. Retrieved from https://levvvel.com/what-is-kernel-level-anti-cheat-software/

[63] Lavrijsen, M. (2021). Disable PatchGuard and DSE at boot time. Retrieved from https://github.com/Mattiwatti/EfiGuard

[64] Lechtik, M. and Dedola, G. (2021). Operation TunnelSnake. Retrieved from https://securelist.com/operation-tunnelsnake-and-moriya-rootkit/101831/

[65] Lechtik, M., Berdnikov, V., Legezo, D., Borisov, I. (2022). MoonBounce: the dark side of UEFI firmware. Kaspersky Lab. Retrieved from https://securelist.com/moonbounce-the-dark-side-of-uefi-firmware/105468/

[66] Lechtik, M., Kayal, A., Rascagneres, P., Berdnikov, V. (2021). GhostEmperor: From ProxyLogon to kernel mode. Kaspersky Lab. Retrieved from https://securelist.com/ghostemperor-from-proxylogon-to-kernel-mode/104407/

[67] Lefferts, R. (2021). Gartner names Microsoft a Leader in the 2021 Endpoint Protection Platforms Magic Quadrant. Retrieved from https://www.microsoft.com/security/blog/2021/05/11/gartner-names-microsoft-a-leader-in-the-2021-endpoint-protection-platforms-magic-quadrant/

[68] Lyk, J. (2020). Easy defender disabler. Retrieved from https://threadreaderapp.com/thread/1339437249528795136.html

[69] Malvica, M. (2020). Silencing the EDR. How to disable process, threads, and image-loading detection callbacks. Retrieved from https://www.matteomalvica.com/blog/2020/07/15/silencing-the-edr/

[70] Maude, J. (2021). TrickBot Attack Chain: Deconstructed & Mitigated. BeyondTrust. Retrieved from Retrieved from https://www.beyondtrust.com/blog/entry/trickbot-attack-chain-deconstructed-mitigated

[71] Menegus, B. (2022). What's the Deal With Anti-Cheat Software in Online Games? Wired. Retrieved from https://www.wired.com/story/kernel-anti-cheat-online-gaming-vulnerabilities/

[72] Microsoft Corporation. (2015). AvScan File System Minifilter Driver. Microsoft Windows Driver Kit (WDK). Retrieved from https://github.com/gochenming/Windows-driver-samples/blob/fb242817317ff9b144fad28f4e3884207d101e35/filesys/miniFilter/avscan/filter/avscan.h

[73] Microsoft. (2021). Improve kernel security with the new Microsoft Vulnerable and Malicious Driver Reporting Center. Retrieved from

https://www.microsoft.com/security/blog/2021/12/08/improve-kernel-security-with-the-new-microsoft-vulnerable-and-malicious-driver-reporting-center/
[74] MITRE. (2021). Exploitation for Privilege Escalation. Retrieved from https://attack.mitre.org/techniques/T1068/
[75] MSDN. (2021a). Privilege Constants (Authorization). Windows App Development. Retrieved from https://docs.microsoft.com/en-us/windows/win32/secauthz/privilege-constants
[76] MS-ISAC. (2020). TrickBot: Not Your Average Hat Trick – A Malware with Multiple Hats. Multi-State Information Sharing and Analysis Center (MS-ISAC), Center for Internet Security (CIS). Retrieved from https://www.cisecurity.org/insights/blog/trickbot-not-your-average-hat-trick-a-malware-with-multiple-hats
[77] MSRC. (2021). Investigating and Mitigating Malicious Drivers. Retrieved from https://msrc-blog.microsoft.com/2021/06/25/investigating-and-mitigating-malicious-drivers/
[78] Naceri, A. (2021). Shutting Down Anti-malware Protection (Part 1) - Windows Defender Antivirus. Retrieved from https://halove23.blogspot.com/2021/08/executing-code-in-context-of-trusted.html
[79] Narib, (2019). Understanding WdBoot (Windows Defender ELAM). Retrieved from https://www.n4r1b.com/posts/2019/11/understanding-wdboot-windows-defender-elam/
[80] Narib, (2020). Dissecting the Windows Defender Driver – WdFilter. Series 1-4. Retrieved from https://www.n4r1b.com/posts/2020/01/dissecting-the-windows-defender-driver-wdfilter-part-1/
[81] Oki, K. (2021a). A PoC for Mhyprot2.sys vulnerable driver that allows read/write memory in kernel/user via unprivileged user process. Retrieved from https://github.com/kkent030315/evil-mhyprot-cli
[82] Oki, K. (2021b) x64 Windows PatchGuard bypass, register process-creation callbacks from unsigned code. Retrieved from https://github.com/kkent030315/NoPatchGuardCallback
[83] Ortiz, B. (2020). Creating AV Resistant Malware — Part 4. Retrieved from https://blog.securityevaluators.com/creating-av-resistant-malware-part-4-6cb2d215a50f
[84] Pérez-Sánchez, A. and Palacios, R. (2022). Evaluation of Local Security Event Management System vs. Standard Antivirus Software. Applied Sciences. 12, 1076, DOI: https://doi.org/10.3390/app12031076
[85] Pham, HD., Nguyen, V.T., Tiep, M.V., Hien, V.T., Huy, P.P., Vuong, P.T. (2021). Evading Security Products for Credential Dumping Through Exploiting Vulnerable Driver in Windows Operating Systems. In: Dang, T.K., Küng, J., Chung, T.M., Takizawa, M. (eds) Future Data and Security Engineering. Big Data, Security and Privacy, Smart City and Industry 4.0 Applications. FDSE 2021. Communications in Computer and Information Science, vol 1500. Springer, Singapore. https://doi.org/10.1007/978-981-16-8062-5_36
[86] Positive Technologies (2021). Rootkits: evolution and detection methods. Retrieved from https://www.ptsecurity.com/upload/corporate/ww-en/analytics/PT_Rootkit_ENG.pdf
[87] Poslušný, M. (2022). Signed kernel drivers – Unguarded gateway to Windows' core. Retrieved from https://www.welivesecurity.com/2022/01/11/signed-kernel-drivers-unguarded-gateway-windows-core/
[88] purpl3f0x (2021). Bypassing Defender on modern Windows 10 systems. Retrieved from https://www.purpl3f0xsecur1ty.tech/2021/03/30/av_evasion.html
[89] pwn1sher. (2022). A small POC to make defender useless by removing its token privileges and lowering the token integrity. Retrieved from https://github.com/pwn1sher/KillDefender
[90] Riley, S. (2006). WCI 442 Windows Vista System Integrity Technologies. Microsoft Corporation. Retrieved from https://download.microsoft.com/download/f/0/b/f0b67ace-f17d-4306-bde2-91abbee80304/WCI442.pdf
[91] Rui., R. (2020). Windows Kernel Ps Callbacks Experiments. Retrieved from http://blog.deniable.org/posts/windows-callbacks/

[92] Sandker, C. (2018). A Windows Authorization Guide. https://csandker.io/2018/06/14/AWindowsAuthorizationGuide.html

[93] Sde-Or, R., Voronovitch, E. (2022). New Milestones for Deep Panda: Log4Shell and Digitally Signed Fire Chili Rootkits. Fortinet. Retrieved from https://www.fortinet.com/blog/threat-research/deep-panda-log4shell-fire-chili-rootkits

[94] Secarma, (2021). Bypassing Windows Defender with Environmental Decryption Keys. Retrieved from https://www.secarma.com/bypassing-windows-defender-with-environmental-decryption-keys/

[95] Shafir, Y. (2022). HyperGuard – Secure Kernel Patch Guard: Part 1 – SKPG Initialization. Winsider Seminars & Solutions Inc. Retrieved from https://windows-internals.com/hyperguard-secure-kernel-patch-guard-part-1-skpg-initialization/

[96] Shaw, J. (2021). Process Herpaderping – Windows Defender Evasion. Retrieved from https://pentestlaboratories.com/2021/01/18/process-herpaderping-windows-defender-evasion/

[97] Shoeb, A. (2021). Dark Web: A Haven for Fake Digital Certificates. Security Boulevard. Retrieved from https://securityboulevard.com/2021/08/dark-web-a-haven-for-fake-digital-certificates/

[98] Sordum. (2021). Defender Control v2.0. Retrieved from https://www.sordum.org/9480/defender-control-v2-0/

[99] Spinney, J. (2019). Behind Enemy Lines: A Pen Tester's Take on Evading AMSI. Retrieved from https://www.wolfandco.com/resources/insights/behind-enemy-lines-a-pen-testers-take-on-evading-amsi/

[100] Stein, Z. (2020). Blinding EDR On Windows. Retrieved from https://synzack.github.io/Blinding-EDR-On-Windows/

[101] Szeles, J, G. (2021). Debugging MosaicLoader, One Step at a Time. Retrieved from https://www.bitdefender.com/files/News/CaseStudies/study/400/Bitdefender-PR-Whitepaper-MosaicLoader-creat5540-en-EN.pdf

[102] T0mux (2018). Evading Windows Defender behavior detection with process injection. Retrieved from https://www.opencyber.com/evading-windows-defender/

[103] Teodorescu, C., Korkin, I., and Golchikov A., (2021). Veni, No Vidi, No Vici: Attacks on ETW Blind EDR Sensors. Black Hat Europe Conference. Retrieved from https://www.blackhat.com/eu-21/briefings/schedule/#veni-no-vidi-no-vici-attacks-on-etw-blind-edr-sensors-24842

[104] Thompson, C. (2017). ATP runs as "Protected Process Light" and "Not_Stoppable". Retrieved from https://twitter.com/retbandit/status/901477187022233600

[105] TNP. (2022). Nerftoken implemented in Golang. Retrieved from https://github.com/tnpitsecurity/nerftoken-go?ref=golangexample.com

[106] Tung, L. (2019). Top Windows Defender expert: These are the threats security hasn't yet solved. ZDNet. Retrieved from https://www.zdnet.com/article/top-windows-defender-expert-these-are-the-threats-security-hasnt-yet-solved/

[107] Unknow101. (2022). A simple python packer to easily bypass Windows Defender. Retrieved from https://github.com/Unknow101/FuckThatPacker

[108] Vella, C. (2019). Reversing & bypassing EDRs. CrikeyCon 2019. Retrieved from https://www.youtube.com/watch?v=85H4RvPGIX4&ab_channel=CrikeyCon

[109] Vijayan, J. (2021). Microsoft-Signed Rootkit Targets Gaming Environments in China. Retrieved from https://www.darkreading.com/attacks-breaches/microsoft-signed-rootkit-targets-gaming-environments-in-china

[110] VL. (2020). Removing Process Creation Kernel Callbacks. Retrieved from https://medium.com/@VL1729_JustAT3ch/removing-process-creation-kernel-callbacks-c5636f5c849f

[111] Xqrzd. (2015). Hazard Shield. Retrieved from

https://github.com/xqrzd/HazardShield/blob/master/HzrFilter/Protect.c

[112] Yasser (2019). A tool to kill antimalware protected processes. Retrieved from https://github.com/Yaxser/Backstab

[113] Yosifovich, P., Russinovich, M., Solomon, D., Ionescu, A. (2017). Windows Internals, Part 1: System architecture, processes, threads, memory management, and more (Developer Reference) 7th Edition. Microsoft Press. Redmond, Washington.

[114] Wang, J.P., Zhao, P., & Ma, H.T. (2017, July). HACS: A Hypervisor-Based Access Control Strategy to Protect Security-Critical Kernel Data. 2nd International Conference on Computer Science and Technology (CST 2017). Guilin, China, DOI: https://doi.org/10.12783/dtcse/cst2017/12516

[115] Korkin, I. (2018-a). Hypervisor-Based Active Data Protection for Integrity and Confidentiality of Dynamically Allocated Memory in Windows Kernel. Paper presented at the Proceedings of the 13th Annual Conference on Digital Forensics, Security and Law (CDFSL), University of Texas at San Antonio (UTSA), San Antonio, Texas, USA. pp. 7-38 Retrieved from https://igorkorkin.blogspot.com/2018/03/hypervisor-based-active-data-protection.html

[116] Tian, D., Xiong, X., Hu, C., & Liu, P. (2018, March 8). A Policy-Centric Approach to Protecting OS Kernel from Vulnerable LKMs. Journal of Software: Practice and Experience. DOI: https://doi.org/10.1002/spe.2576

[117] Hua, Z., Du, D., Xia Y., Chen H., & Zang, B. (2018). EPTI: Efficient Defence against Meltdown Attack for Unpatched VMs. In Proceedings of the USENIX Annual Technical Conference (ATC). Boston, MA. pp. 255-266. Retrieved from https://www.usenix.org/conference/atc18/presentation/hua

[118] Tian, D., Jia, X., Chen, J., & Hu. C. (2017). An Online Approach for Kernel-level Keylogger Detection and Defense. Journal of Information Science and Engineering. Volume 2. Number 2. pp. 445-461. Retrieved from https://journal.iis.sinica.edu.tw/paper/1/151104-2.pdf?cd=73514F4E04688D3DE